\documentclass[12pt]{article}
\pdfoutput=1

\usepackage{amsmath,amssymb,amscd}

\usepackage{listings}
\usepackage{caption}
\usepackage{dsfont}
\usepackage{slashed}
\usepackage{color}
\usepackage{ulem}
\usepackage[pdftex]{graphicx}
\usepackage{epstopdf}
\usepackage{subfigure}
\usepackage{epsfig}
\usepackage{listings}
\usepackage{caption}
\usepackage{cite}
\usepackage{multirow}
\usepackage{comment}

\newcommand{\beq}{\begin{equation}}
\newcommand{\eeq}{\end{equation}}
\newcommand{\nbea}{\begin{align*}}
\newcommand{\neea}{\end{align*}}
\newcommand{\nbeq}{\begin{equation*}}
\newcommand{\neeq}{\end{equation*}}

\newcommand{\hinf}{H_{\rm inf}}
\newcommand{\hinfinv}{H_{\rm inf}^{-1}}
\newcommand{\rfh}{Q_{\rm F\hinf}}
\newcommand{\pgw}{P_{\rm gw}}
\newcommand{\ogw}{\Omega_{\rm gw}}
\newcommand{\rgw}{\rho_{\rm gw}}
\newcommand{\teq}{t_{\rm eq}}

\usepackage{multirow}
\usepackage{array}
\newcolumntype{M}[1]{>{\centering\arraybackslash}m{#1}}
\newcolumntype{N}{@{}m{0pt}@{}}

\begin{document}

\baselineskip=21pt

\begin{center}

{\large {\bf Looking at the NANOGrav Signal Through the Anthropic Window \\ of Axion-Like Particles}}

\vskip 0.2in
\vskip 0.2in

{\bf Alexander S.~Sakharov}\textsuperscript{a,b},~
{\bf Yury N.~Eroshenko}\textsuperscript{c} and~
{\bf Sergey G.~Rubin}\textsuperscript{d,e} ~\\

\vskip 0.2in
\vskip 0.2in

{\small {\it

\textsuperscript{a}Physics Department, Manhattan College\\
{\mbox 4513 Manhattan College Parkway, Riverdale, NY 10471, United States of America}\\
\vspace{0.25cm}
\textsuperscript{b}{\mbox Experimental Physics Department, CERN, CH-1211 Gen\`eve 23, Switzerland}\\
\vspace{0.25cm}
\textsuperscript{c}{\mbox Institute for Nuclear Research of the Russian Academy of Science}\\
\mbox{117312 Moscow, Russia}\\
\vspace{0.25cm}
\textsuperscript{d}{\mbox National Research Nuclear University MEPhI (Moscow Engineering Physics Institute)}\\
\mbox{Kashirskoe Shosse 31, 115409 Moscow, Russia}}\\
\vspace{0.25cm}
\textsuperscript{e}{\mbox N.~I.~Lobachevsky Institute of Mathematics and Mechanics, Kazan  Federal  University\\
420008, \mbox{Kremlevskaya  street  18,  Kazan,  Russia}}
\vspace{0.25cm}

}

\vskip 0.2in

{\bf Abstract}

\end{center}

\baselineskip=18pt
\noindent


We explore the inflationary dynamics leading to formation of closed domain walls in course of
evolution of an axion like particle (ALP) field whose Peccei-Quinn-like phase transition occurred
well before inflationary epoch. Evolving after inflation, the domain walls may leave their
imprint in stochastic gravitational waves background, in the frequency range accessible for
the pulsar timing array measurements. We derive the characteristic strain power spectrum
produced by the distribution of the closed domain walls and relate it with the recently
reported NANOGrav signal excess. We found that the slope of the frequency dependence of the
characteristic strain spectrum generated by the domain walls
is very well centered inside the range of the slopes in the signal reported by the NANOGrav.
Analyzing the inflationary dynamics of the ALP field, in consistency with the isocurvature
constraint, we revealed those combinations of the parameters where the signal
from the inflationary induced ALPs domain walls could saturate the amplitude of the NANOGrav excess.
The evolution of big enough closed domain walls may incur in formation of wormholes with the
walls escaping into baby universes. We studied the conditions, when closed walls escaped
into baby universes could leave a detectable imprint in the stochastic gravitational waves background.

\begin{center}
{\it We dedicate this paper to the memory of Roberto Peccei, one of the creators of axion}
\end{center}

\vskip 3mm

Wormholes; Baby universe
\vskip 5mm

\leftline{26 June 2021}

\vskip 10mm


\section{Introduction}
\label{intro}

By their seminal papers~\cite{Peccei,PecceiPRD}, offering the most attractive
solution of the strong CP problem in QCD, Roberto Peccei and Helen Quinn, in fact,
gave rise to an extensive line of actively developing
researches, the axion cosmology~\cite{miss1,miss2,miss3,axionKim1,alp1,axionRev1,axionRev2},
generalized, in the mean time, in
the cosmology of axion-like particles (ALPs)~\cite{alp2,alp3,alp4,axionUnif1,axionUnif2}.

The solution~\cite{Peccei} based on the assumption
of the existence of a global ${\rm U(1)_{\rm PQ}}$ Peccei-Quinn (PQ) symmetry
that is broken, at scale $F$, to a discrete subgroup $Z_N$, which generates
a pseudo-Nambu-Goldstone (PNG) boson, known as axion~\cite{axionKim1}.
Due to the QCD anomaly the axion obtains a periodic potential
with N distinct minima and hence acquire a mass specifically
connected with the PQ scale as $\propto \Lambda_{\rm QCD}^2/F$.
Such axion has been baptized the QCD axion~\cite{axionKim1,alp1,axionRev1,axionRev2}.
It is suitable to set the PQ symmetry breaking
scale much above the electroweak scale, so that the axion should have a
very reduced coupling to the Standard Model particles, which makes the axion ``invisible''
rendering it a very promising candidate to act
as dark matter (DM)~\cite{axionKim1,alp1,axionRev1,axionRev2,axionWallStr1}.

The axion may naturally realized in string
theory~\cite{alpString1,alpString2,alp2}
where, moreover, the compactification
always generates PQ-like symmetries
which generically produce the multitude of axions
and ALPs. The number and properties of
the ALPs are defined by the specific model
being considered. However, typically those
string models which incorporate the QCD axion
produce ALPs with PQ-like scale of about or higher
than the GUT scale and masses similar to that of the
QCD axion~\cite{alp2}.

In analogy to the QCD axion, a high PQ-like symmetry breaking
scale assumes that this symmetry is broken before inflation.
Thus, during inflation, the ALP behaves as a massless spectator field, which
is made uniform trough the scale of the observable universe. The value
of the ALP field, over this biggest scale, is defined by some random displacement $\theta_0$
from the minimum of its potential, which is called
the misalignment angle.
The misalignment angle $\theta_0$ determines the amplitude of
the ALP field, once its mass exceeds the Hubble rate and
the friction term does non prevent it anymore from oscillations
about its minima. These coherent oscillations of the Bose-Einstein
condensate are pressureless, representing a dust like substance
acting as a cold DM (CDM) in the Universe~\cite{miss1,miss2,miss3}.

Either the QCD axion~\cite{alp1} or ALPs~\cite{alp1,alp3,alp4} with above GUT scale
of the PQ symmetry breaking might be accomodated only if we inflated from a rare patch
of the space with very small misalignment angle $\theta_0<<1$.
The unlikely value can be explained by involving of anthropic
selection based on the arguments that a high axion or ALP density
should be hostile to the development of
life~\cite{antrop1,antrop2,antrop3,antrop4,antrop5}.
In this case, it is said that the axion (or ALP) parameters are located in the
anthropic window.

A massless inflationary ALP spectator acquires the quantum fluctuations
imprinted by the de Sitter expansion. When the ALP obtains its mass, well
after the end of inflation, these fluctuations, being converted into isocurvature
fluctuations, become dynamically relevant. The isocurvature fluctuations
are uncorrelated with the adiabatic fluctuations inherited by
all other matter and radiation from the fluctuations of the inflaton
field~\cite{antrop3,antrop4,isoALPcurvLyth1,axInf1,isoALPcurv1,isoALPcurv4,isoALPcurv5}.
The main observational effect of the ALP isocurvature fluctuations
consists in modification of the Sachs-Wolfe plateau
at small cosmic microwave background (CMB) multipole orders corresponding
to the scales comparable to the size of the last scattering surface.
 So far, there is no trace of the isocurvature fluctuations in existing
 CMB data~\cite{plankX}. Therefore, it would be interesting to study
 opportunities, if any, that ALPs from anthropic window could
 show up in other types of observations.

 The inflationary quantum fluctuations~\cite{infl0,infl1} of the ALP field should lead
 to substantial deviations from the initial misalignment angle~\cite{axInf2},
 at the late stages of the inflation, when scales much smaller
 than that ones relevant for CMB observations were exiting the
 inflating Hubble patches. The deviations can be so large
 that, in some domains, the ALP field is already localized in a
 position to start oscillations about its different minimum,
 once it is liberated from the Hubble friction. In this case
 the neighboring domain with ALP oscillating about different
 minima must be interpolated by a closed domain wall, with stress energy
 density defined by the scale of PQ-like symmetry breaking and
 the mass of the ALP~\cite{we1,we2}. The abundance of such walls of a given size
 is defined by the initial misalignment angle $\theta_0$ and
 the ratio of the PQ-like scale to the inflationary Hubble rate.
 Since, here, we are talking about scales much smaller than
 those ones relevant for the CMB observations, the inhomogeneities,
 caused by dramatic difference in values of the initial amplitudes
 of the ALP oscillations, cannot be observed as the
 isocurvature fluctuations, discussed above. Depending
 on its initial size, the evolution
 of a closed domain wall in the FRW Universe consists in
 either its collapse into a
 primordial black hole (PBH)~\cite{genPBH1,genPBH2,genPBH23,genPBH3,genPBH4,genPBH5,Caretal20,genPBH6}
 or formation of a wormhole with the wall escaping into
 a baby universe.

In both cases of wall's evolution, a certain mass
is set in a motion, which may lead to a production of
gravitational waves (GWs).
Indeed, a collapsing domain should release its asphericity in course of
its entering under the Hubble radius, so that the
biggest wall's fragments make one or few oscillations radiating their energy
in form of GWs with characteristic frequency of about the Hubble rate.
A closed wall forming a wormhole, due to its negative pressure,
may induce sound waves in the interior bulk fluid leading also to
emission of GWs in similar frequency band.

Recently, the North American Nanohertz Observatory for Gravitational Waves
(NANOGrav) collaboration has reported the signal from a stochastic
process from the analysis of 12.5 years data of 45 pulsars
timing array (PTA) measurements~\cite{NANOGrav1}. The signal can
be interpreted as a stochastic GWs
background~\footnote{PTA measurements for detection of
very long gravitational waves have been proposed in~\cite{pta1,pta2}.}
at frequency around
31~nHz. In the mean time, it is still not clear
whether the detected signal is of truly origin from
the stochastic GWs background because of the tension
with previous PTA measurements in similar
frequency range and also there has been not
detected the quadrupole correlations~\footnote{The quadrupol
correlations is a smoking gun of stochastic GWs background
in PTA measurements~\cite{PTAQ1}.}
as yet.

Nevertheless, it was quickly realized, that apart the common astrophysical sources,
such as super massive black holes binaries  (SMBHBs),
the stochastic GWs interpreted signal of the NANOGrav~\cite{NANOGrav1},
might be an aftersound of the vacuum rebuilding processes
which could take place in the early Universe. These are the formation
of cosmic strings~\cite{stringNANO1,stringNANO2,stringNANO3,stringNANO4,stringNANO5},
first order phase transitions~\cite{ptNANO1,ptNANO2},
the formation of domain walls~\cite{wallNANO1,wallNANO2},
turbulent motions occurring at QCD phase transitions~\cite{qcdNANO1,qcdNANO2}
and cosmological inflation~\cite{nanoInfl1}.

In this paper we estimate the spectra of GWs produced
by closed domain walls, the formation of
which is induced in course of inlationary dynamics
of ALPs with parameters,  particularly but not exclusively,
related to the anthropic window. We found that
the walls collapsing into PBHs, at the stage of
occurring of sphericity, can generate the
stochastic GWs background with the characteristic strain power spectrum slope
remarkably well centered within the range of slopes reported in
the NANOGrav signal. Analyzing the inflationary dynamics
of the ALP field, in consistency with the isocurvature
constraint, we reviled such combinations of the
parameters where the signal from the inflationary induced ALPs walls
could saturate the amplitude of the NANOGrav excess.
As a by product of the analyzes, we have elucidated the
conditions, when closed domain walls escaped into baby
universes could leave a detectable imprint in
the stochastic gravitational waves background.

The rest of the paper is organized as follows. In Section~\ref{size}, we describe
the mechanism of origin of closed domain walls in presence of ALP field during inflation
and their subsequent evolution. In Section~\ref{stochGW} and Section~\ref{GWbaby}
we derive the spectra of stochastic GWs background generated by collapsing and
escaping domain walls, respectively. In Section~\ref{nanoAxion},  the derived spectra are related to
the spectral slope and the amplitude of the signal reported by the NANOGrav.
In Section~\ref{infALP}, we analyze the parameters of ALP in the light its inflationary dynamics and
saturation of the NANOGrav signal amplitude. In Section~\ref{isocurv}, we define
the combinations of parameters in scenarios where ALPs could manifest themselves
in the NANOGrav signal, being compatible with DM density and isocurvature constraints.
In Section~\ref{sec:bu}, we discuss possible signatures of domain walls escaping into baby universes
in stochastic GWs background signals. In Section~\ref{sec:bh}, we check the consistency of the PBHs production
by closed domain walls versus existing PBHs constraints. We conclude in Section~\ref{sec:conclusions}.

\section{Inducing domain walls formation in presence of ALP while inflating}
\label{size}

We consider potential of a complex scalar field
$\phi=r_{\phi}e^{i\theta}$ with self-interaction constant $\lambda$
\beq
\label{PQ1}
	V(\phi)=\lambda\left(|\phi|^2-F^2/2\right)^2,
\eeq
whose ${\rm U(1)_{PQ}}$ symmetry is broken at the scale obeying the
condition
\beq
\label{PQF}
\rfh\equiv\frac{F}{\hinf}>(2\pi)^{-1},
\eeq
which implies that the radial mass $m_r=\sqrt{\lambda}F$ exceeds the inflationary
Hubble rate $\hinf$, so that the field $\phi$ resides in the ground state
during inflation.
The ALP is then the pseudo Nambu-Goldtone (PNG) boson of the broken ${\rm U(1)_{PQ}}$
symmetry
\beq
\label{png1}
\theta ({\bf x})=a({\bf x})/F.
\eeq
Unlike in case of the QCD axion, the mass of ALP field $a$ should not have specific
connection to the scale of the ${\rm U(1)_{PQ}}$ symmetry breaking $F$. In case
of realization of axion scenario in string theory, compactifications always
generate PQ like symmetries~\cite{alpString1,alpString2,alp2} and generically
produce a multitude of ALPs.

We assume that
ALP potential, being generated by some exotic, strongly interacting, sector,
may be approximated as
\beq
\label{axion1}
V(a )\approx\Lambda^4\left[1-\cos\left(\frac{a}{F}\right)\right],
\eeq
where $\Lambda$ is a scale set by the ALP coupling to
instantons and may span a wide range from QCD scale to the string scale,
with each ALP associated with its own gauge group~\cite{alpString1}.
The ALP's mass reads
\beq
\label{Amass1}
m_{\theta}^2=\frac{\partial^2V}{\partial a^2}=\frac{\Lambda^4}{F^2},
\eeq
so that the ALP is described by two out of the three parameters $F$,
$m_{\theta}$ and $\Lambda$.

If string theory can produce a QCD axion with large scale of PQ symmetry
breaking and hence with small mass, one can expect that the spectrum of
ALPs should include masses within few orders of magnitude of that one
of the QCD axion. String theory models, as for example are described
in~\cite{alpString1,alpString2,alp2}, can accommodate many ALPs
with $F$ values about the GUT scale. The masses of such ALPs
should be homogeneously distributed on a log scale~\cite{alpString1}, with
several ALPs per energy decade.

During inflation and some period of Friedman–Robertson–Walker (FRW) epoch, the energy density
(\ref{axion1}) is negligible starting playing a significant role in the dynamics
of phase $\theta ({\bf x})$ at the moment when the mass of the ALP (\ref{Amass1}) overcomes the Hubble rate.
The potential (\ref{axion1}) possess a discrete
set of degenerate minima corresponding to the phase values
$\theta_{\rm min}=0,\pm 2\pi ,\pm 4\pi\dots$.
After the inflation ended, as soon as the $m_{\theta}$
started to exceed the Hubble rate, the field
$\theta ({\bf x})$ becomes oscillating about
the minima, so that the energy stored in the potential (\ref{axion1}) gets converted
via misalignment angle mechanism~\cite{miss1,miss2,miss3}
into the ALPs in form of Bose-Einstein condensate behaving
as a cosmological cold DM (CDM)~\footnote{If the $m_{\theta}$ is very large,
which is to say that $\Lambda\gtrsim\hinf$, the ALP begin oscillating before the inflation,
so that the Bose-Einstein condensate is inflated away.}
(see discussion in Section~\ref{isocurv}).

At the condition of vanishing $m_{\theta}$, the inflationary dynamics of the phase is driven by
the quantum fluctuations of magnitude~\cite{antrop1,axInf1,axInf2,isoALPcurv1}
\beq
\label{step1}
\delta\theta\approx\frac{1}{2\pi}\rfh^{-1},
\eeq
taking place each Hubble time $\hinf^{-1}$.
The amplitude of these fluctuation freezes out due to a large
friction term in the equation of motion of the massless PNG spectating
the de Sitter background, whereas their wavelength grows
exponentially. Such process resembles
one dimensional Brownian motion of variable $\theta$,
inducing, at each inflationary e-fold, a classical increment/decrement
of the phase by factor $\delta\theta$ given in (\ref{step1}).
Therefore, once the inflation began in a single, causally
connected domain of the horizon size $\hinf^{-1}$, being filled with initial phase
value $\theta_0<\pi$, it progressed producing exponentially growing
domains with phase values being more and more separated
from the initial phase value $\theta_0$. It is unavoidable, that in some
domains the phase values has grown above $\pi$, so that one expects to see
a Swiss cheese like picture where the domains with
phase value $\theta >\pi$ are inserted into a space remaining filled in with  $\theta <\pi$.
Therefore, in the domains where the inflationary dynamics has induced
the phase value $\theta >\pi$ the oscillations of the ALP field will occur
about the minimum in $\theta_{\rm min}= 2\pi$, while in the surrounding space
with $\theta <\pi$ the oscillations will chose the minimum in  $\theta_{\rm min}= 0$.
It is well known~\cite{wall1}, that in such a setup, in case of periodic potential (\ref{axion1}),
two above vaccua should be interpolated by the sine-Gordon kink solution
\beq
\label{sinG1}
a(z)=4F\tan^{-1}\exp (m_{\theta}z)
\eeq
interpreted as a domain wall of width $m_{\theta}^{-1}$ posed perpendicularly to z axis, with the stress energy density
\beq
\label{sigma1}
\sigma =4\Lambda^2F = 4m_{\theta}F^2.
\eeq
Therefore, one expects that every boundary separating a domain with phase value
$\theta >\pi$ from the space of $\theta <\pi$ traces the surface of a closed
domain wall to be formed in some time after the end of inflation, when $\theta ({\bf x})$
starts to oscillate. Since after inflation, during FRW epoch, the wall of size
$R(t_{\rm end})$ is simply conformally stretched by the expansion of the universe
\beq
\label{exp1}
R(t)=\frac{a(t)}{a(t_{\rm end})}R(t_{\rm end}),
\eeq
where $a(t)$ is the scale factor and $t_{\rm end}$ corresponds to the end of the inflationary epoch,
the size distribution of such a sort of inflationary induced walls' seeding contours
can be directly mapped into the size distribution of the domain walls. In what follows, we derive
this size distribution.

A domain wall's seeding contour of radius $R\approx\hinfinv$ emerging during inflation of total duration $t_{\rm inf}$ at the time moment
$t_s$, when the universe is still have to inflate over $\Delta N_s=\hinf (t_{\rm inf}-t_s)=N_{\rm inf}-N_s$ e-folds, is
getting stretched in course of the expansion as
\beq
\label{R1}
R(\Delta N_s)\approx\hinfinv e^{\Delta N_s}.
\eeq
The number of contours created in a comoving volume $d{\bf V}$ within e-fold interval
$dN_s$ is given by
\beq
\label{dN1}
dN=\Gamma_s\hinf^3e^{3N_s}d{\bf V}dN_s,
\eeq
where $\Gamma_s$ is the contours' formation rate per Hubble time-space volume $\hinf^{-4}$, which, in general,
should depend on formation instant $N_s$ and is defined by the inflationary
dynamics of the ALP spectator field (see the detailed discussion in Section \ref{infALP}).
Expressing $N_s$ from (\ref{R1}), one can write down
the number distribution of seeding contours with respect to their physical radius $R$
\beq
\label{dN2}
d{\cal N} = \Gamma_s\frac{e^{3N_s}d{\bf V}}{R^4}dR.
\eeq
Thus, the number density in physical inflationary volume $dV_{\rm inf}=e^{3N_s}d{\bf V}$
reads
\beq
\label{n1}
\frac{dn}{dR}=\frac{d{\cal N}}{dRdV_{\rm inf}}=\frac{\Gamma_s}{R^4}.
\eeq
At the end of inflation the distribution (\ref{n1}) spans the range of scales from $R_{min}\simeq\hinfinv$
to $R_{max}\equiv R(\Delta N_{\pi})\approx\hinfinv e^{(N_{\rm inf}-N_\pi)}$, where $N{_\pi}$ is the number of e-folds when the angular degree
of freedom $\theta $ overcame $\pi$, since the beginning of inflation lasting enough
$N_{\rm inf}$ e-folds needed to solve horizon and
flatness problem~\footnote{A different mechanism of production of topological defects, at the inflationary stage,
leading to similar size distribution, has been considered in~\cite{wallPBH1,wallPBH2,vilenkin1}.}.

Once the phase oscillations begin and a domain wall is materialized, replacing the surface of its seeding contour,
it will stay essentially at rest relative to the Hubble flow until its size becomes comparable
with the Hubble radius  $R(t_H)\approx H^{-1}$ defined by the Hubble rate $H=\dot a/a$.
Evidently, at the entering into the horizon the surface tension
forces developed in the wall tend to minimize its surface. The dynamics of spherical domain wall
has been studied in~\cite{wallEvol1} in asymptotically flat space. It was shown, that in this case
the internal domain wall metric is Minkowski while the external one is Schwartzshild, so that
the wall being initially bigger than its Schwartzshild radius always collapses
into a BH singularity. Thus, the ALP field induced domain wall should obtain a spherical geometry and contract toward the
center. The regime (\ref{PQF}) assumes that the ALP's coupling to the Standard Model particles is strongly suppressed
by the large scale $F$, so that the ALP wall interacts very weakly with matter and hence nothing can prevent
the wall finally to be localized within its gravitational radius and hereby deposit its energy into a
PBH~\cite{we1,we2}.

A more general case of a false vacuum bubble surrounded by true vacuum,
and the study the motion of the domain wall at
the boundary of these two regions are presented in~\cite{wallEvol2,wallEvol3,wallEvol4}.
In particularly, it was shown that if the energy density
inside the bubble is vanished and the unique source of gravitation field is the domain wall,
as for example occurs when the domain wall emerges from
a "white hole" singularity, it will expand unbounded to escape in a baby universe.
The baby universe is initially connected to the asymptotically flat patch of space
by a wormhole, which eventually is getting pinched off leading to changes in topology.
A similar picture of evolution may take place in case of a wall
exceeding by its mass the energy of fluid inside the Hubble radius, at the entering.
Indeed, as it is described
in~\cite{wallPBH1,wallPBH2}, due to its repealing nature,
the wall, with dominating gravitational effects, pushes away
the radiation/matter fluid, creating rarefied layers in the vicinity
of its both interior and exterior surfaces. Since the exterior FRW Universe
continues to follow the Hubble expansion, while the wall extends
exponentially in proper time, it has to create a wormhole, through
which it escapes into a baby universe. Finally, the wormhole is getting
pinched off, within about its light crossing time, so that
observers on either side of the throat of the wormhole
is seeing a BH forming, probably having unequal masses on both sides.

Here we notice that, in general, along with the phase fluctuations
(\ref{step1}), one has to expect the radial fluctuations which
might kick the radial component $r_{\phi}$ out of its vacuum
state $r_{\phi}=F$, over the top of the Mexican hat potential (\ref{PQ1}).
This would lead to the formation of strings along the lines
in space where $r_{\phi}=0$, so that one arrived to a realization
of the scenario of the system of walls bounded by strings~\cite{wallStr1,wall1}.
The radial fluctuations, in the context of string formation
during inflation have been investigated both, analytically and numerically
in~\cite{axInf2}. In particular, it was elucidated that the strings
would be formed if the Hubble rate were exceeding the radius of the
Mexican hat potential. More precisely, if the condition (\ref{PQF})
is violated, during inflation~\footnote{We might assume that the Hubble rate
was very high, well before $N_{\rm end}$. However, in this case, all
inflationary formed strings would be washed out from our Hubble volume.}.
In some sense, there is a "Hawking like" temperature of order of
$\hinf$ which takes over to drive a thermal like phase transition
with ALP string formation. In contrary, $r_{\phi}$
sits in its vacuum $r_{\phi}=F$ whenever (\ref{PQF}) stays in tact,
for a reasonable value of $\lambda\simeq 10^{-2}$ and simple
inflationary model~\cite{axInf2}. Therefore, in the regime
$\rfh >1$ relevant for the analysis presented bellow, only
closed domain walls can be formed, since the formation of
strings is prevented.

Actually, the closed walls can decay
due to quantum nucleation of holes bounded by strings
on their surfaces~\cite{wallStr2,wall1}. The process of the hole
nucleation can be described semi-classically by an instanton
so that the decay probability is expressed as~\cite{wallStr2}
\beq
\label{instanton1}
P_{\rm hole}\simeq A\exp\left(-\frac{16\pi\mu^3}{3\sigma^2}\right),
\eeq
where $\mu\simeq\pi f^2$ is the string tension and $A$ is a non
infinitesimally small factor
which can be calculated from the analysis of small perturbations
around the instanton solution. Thus, applying (\ref{sigma1}),
one can see that the probability (\ref{instanton1}) is suppressed
by factor $\simeq\exp (-10^2(F/\Lambda )^4)$, which is very small
for a large hierarchy of the scales, $F/\Lambda >>1$, which is the
case for the parameters of ALPs, discussed bellow. Thus, one can
affirm that the closed ALP's domain walls considered here are
stable relative to the quantum nucleation of holes bounded
by strings.

The process of occurring of the spherical shape of a collapsing closed domain wall
should be accompanied by a production of
gravitational waves~\cite{wallGW0,wallGW1,wallGW2}. Also, one might assume that sound waves
induced in the interior radiation fluid by walls escaping into wormholes may also
contribute as a source of stochastic gravitational waves background.

Bellow, we study the domain walls formed as a result of the dynamics
of ALP field found itself in the regime of the inflationary spectator,
as a source of primordial stochastic GWs background, in both collapsing and expanding regimes
outlined above. We derive the spectra of the GWs generated by domain walls of sizes
distributed with respect to (\ref{n1}). The spectra, being related to the NANOGrav
signal~\cite{NANOGrav1}, elucidate the detectability of the evidence of existing of ALPs, including
those characterized by parameters relevant for the anthropic window.

\section{Stochastic gravitational waves background generated by the collapsing ALP's walls}
\label{stochGW}
The ceeding contours of closed ALP's domain walls, formed in the way described above,
have complex structure which is characterized by inflections (or folds) in all possible scales
including those comparable with the size of the walls as whole objects
(see a detailed discussion in~\cite{genPBH6}).
After the materialization, as soon as a folded fragment of a closed domain wall becomes causally connected, namely
it finds itself within a Hubble horizon comparable to the scale of the inflection's curvature, this fragment is getting
stretched due to the wall's tension forces, so that its shape is smoothed out within the Horizon scale.
Therefore, once a close domain wall finds itself as whole causally connected unit, being localized
within a Hubble horizon which encompasses it as complete object, the wall will mostly contains fragments
inflected (folded) in scales which are as big as the scale of the horizon.
The stretching of these biggest fragments leads to oscillations in scales comparable
with the size of the whole closed wall. Since the ALP's walls, for the PQ-like scale of
symmetry breaking considered here, are coupled extremely weekly with the SM particles media
the oscillations can be dumped releasing their energy in form of
GWs contributing into the stochastic GWs background, which might be observed today.
Finally the closed domain wall obtains a spherical shape  in course
of its one or, may be, few oscillations
within the scale of the Hubble radius~\footnote{The described picture akin to the case
of oscillations of big soap bubbles which being detached from their seeding frame
initially have an irregular shape and start to oscillate at low spherical harmonic multipoles
with amplitude comparable to the size of the whole bubble.}.

Here, we are interested in the presently existing stochastic background of GWs,
created by domain walls described above, as the dimensionless fraction of the critical density expressed by the
energy of GWs in units of logarithmic interval of frequency
\beq
\label{ogw1}
\ogw (\ln f)=\frac{8\pi G}{3H_0^2}f\rgw (t_0,f),
\eeq
where $\rgw$ is the energy density of GWs in unit frequency.
The energy density in comoving volume is just the total energy deposited
in GWs
\beq
\label{rgw0}
\rgw (t_0,f) = \int_{t_{\rm sw}}^{t_0}\frac{dt}{(1+z(t))^4}\pgw (t,f')\frac{\partial f'}{\partial f},
\eeq
where $\pgw (t,f')$ is the total GWs energy power
emitted at time $t$ into unit range of frequencies. The frequency $f$ today
had been emitted as frequency $f'=(1+z)f$, so that we can write
\beq
\label{rgw1}
\rgw (t_0,f) = \int_{t_{\rm sw}}^{t_0}\frac{dt}{(1+z(t))^3}\pgw (t,(1+z)f).
\eeq
This background can be expressed as its
power spectral density~\cite{gwexp1}
\beq
\label{psd1}
S_h(f)=\frac{3H_0^2}{2\pi^2f^3}\ogw (\ln f).
\eeq
The pulsar timing arrays like NANOGrav~\cite{NANOGrav1} use so called
characteristic strain
\beq
\label{h1}
h_c=(fS_h(f))^{1/2}.
\eeq

The power of the
emission of the GWs by an individual domain wall
can be estimated using the formula of quadrupole radiation power by a massive object~\cite{Landau_V2}
\beq
\label{pw1}
{\pgw}_i\approx \frac{G}{C_2}\left(\frac{d^3I}{dt^3}\right)^2,
\eeq
where $I$ is the quadrupole moment of the object and $C_2=45$. We assume that just in the course of a complete
entering of a closed wall of size $R$ into the matchable Hubble radius, its biggest fragment(s)
pass through one or few oscillations within the horizon scale. For such a fragment of a domain wall the
quadrupole moment is estimated as
\beq
\label{qm1}
I\simeq\sigma H^{-4}\simeq\sigma f'^{-4},
\eeq
where $f'\approx R^{-1}$ is the frequency of the GW emitted within a Hubble
horizon of size $R$.
Therefore, in this case, the power (\ref{pw1}) can be expressed as
\beq
\label{pw2}
{\pgw}_i\approx \frac{G}{C_2}\frac{\sigma^2}{f'^2}.
\eeq

The GWs power $\pgw (t,f')$ is contributed by domain walls
that collapse having radii between $R$ and $R+dR$ at Hubble crossing. For such
domain walls, collapsing before mater-radiation equality $t_H<\teq$,
the Hubble crossing is $t_H=R/2$, so that their number density evolves as
\beq
\label{n2}
dn(t,f')\simeq\Gamma_s\left(\frac{R}{t}\right)^{3/2}\frac{dR}{R^4}\approx \Gamma_st^{-3/2}f'^{1/2}df',
\eeq
where $t>t_H$.
Thus, one can write
\beq
\label{pw3}
\pgw (t,f') = \frac{dn(t,f')}{df'} {\pgw}_i\approx \Gamma_s\frac{G\sigma^2}{C_2}t^{-3/2}f'^{-3/2}.
\eeq

Substituting (\ref{pw3}) in (\ref{rgw1}) we can change the integration variable using
\beq
\label{int1}
dt = -\frac{dz}{H(z)(1+z)}
\eeq
to get
\beq
\label{rgw2}
\rgw (t_0,f) = \Gamma_s\frac{G\sigma^2}{C_2}f^{-3/2}\int_{0}^{\infty}\frac{t(z)^{-3/2}dz}{H(z)(1+z)^{11/2}}.
\eeq
For the $\Lambda$CDM cosmology we have
\beq
\label{hz1}
H(z)=H_0(\Omega_{\Lambda}+(1+z)^3\Omega_m+{\cal G}(z)(1+z)^4\Omega_{r})^{1/2},
\eeq
where the function ${\cal G}(z)$ accounts for changes of number of relativistic
degrees of freedom at early time, $H_0=100h$~km/s/Mpc. As the standard
parameters we will use $\Omega_{\Lambda}=0.69$, $\Omega_m=0.31$,
$h^2\Omega_r=5\times 10^{-5}$ and $h=0.68$.

We argue in Section \ref{infALP} that the scales, relevant for the dynamics of domain walls
in the current study, emerge well before equality epoch. Thus, one can integrate
(\ref{rgw2}) in the radiation dominated area ignoring the change of the number
of degrees of freedom. The ratio of the current scale factor $a_0$ to that one corresponded
to the equality of matter and radiation $a_{\rm eq}$ occurred at
red shift $z_{\rm eq}$ is given by
\beq
\label{a1}
\frac{a_0}{a_{\rm eq}}=(1+z_{\rm eq})=\frac{\Omega_m}{\Omega_r}\approx 2\times 10^{4}(\Omega_mh^2).
\eeq
In the radiation dominated regime one can write
\beq
\label{hz2}
H(z)=(1+z)^2H_r,
\eeq
so that
\beq
\label{tz1}
t(z)=\frac{1}{2(1+z)^2H_r},
\eeq
where the contribution of radiation to the Hubble rate is given by
\beq
\label{Hr1}
H_r=H_0\Omega_r^{1/2}.
\eeq
Therefore, the integral (\ref{rgw2}) can be reduced to
\beq
\label{rgw3}
\rgw (t_0,f) = \frac{2\sqrt{2}}{C_2}\Gamma_sG\sigma^2f^{-3/2}H_r^{1/2}\int_{0}^{\infty}\frac{dz}{(1+z)^{9/2}}=
\frac{4\sqrt{2}}{7C_2}\Gamma_sG\sigma^2H_0^{1/2}\Omega_r^{1/4}f^{-3/2}.
\eeq
Finally, one can express the fraction (\ref{ogw1}) of the stochastic GWs background generated
by the collapsing domain walls as follows
\beq
\label{ogw2}
\ogw (\ln f) = \frac{32\pi\sqrt{2}}{21C_2}\frac{\Omega_r^{1/4}}{H_0^{3/2}}\Gamma_s(G\sigma)^2f^{-1/2}.
\eeq

\section{Gravitational wave signal induced by walls escaping into baby universes}
\label{GWbaby}

Due to its repulsive nature the domain wall, which dominates
by its mass over the
energy of the interior fluid
at the Hubble radius entering instant, pushes away from its interior surface
the radiation bulk fluid and then inflates out forming a wormhole~\cite{wallPBH1,wallPBH2}. We assume that
the radiation fluid should respond on this act of repulsion by sound waves of characteristic
wavelength $\simeq H^{-1}$ which set in a motion a fraction of the mass
contained inside the Hubble radius. This motion should generate GWs
of frequency $f'\approx H$, which will give a contribution into
stochastic GWs background~\footnote{A similar kind of contribution is usually
considered in connection with GW signal generated by a first order
phase transitions, see for example~\cite{phaseTr1}.}.

In such a setup the quadrupole moment in (\ref{pw1}) can be estimated as
\beq
\label{qmb1}
I_{b}\simeq \kappa M_b(H)H^{-2}\approx\frac{\kappa}{2Gf'^3},
\eeq
where $M_b(H)$ is the mass of the radiation fluid that would be contained
inside a dominating wall at the instant of its entering into the Hubble
radius and $\kappa$ is the fraction of this mass which is set in a motion
by the sound waves. Therefore, the power (\ref{pw1}) generated
by an individual wall can be expressed as
\beq
\label{bpw1}
P_{{gb}_i}\approx \frac{\kappa^2}{4C_2G}.
\eeq
In case of size independent $\kappa$, which is used bellow,
the power (\ref{bpw1}) is frequency independent as well.
For the domain walls, escaping before mater-radiation equality $t_H<\teq$,
and hence pushing away the interior radiation fluid inside the Hubble
radius $t_H=R/2$, the differential power can be estimated in the analogy to
(\ref{pw3}), and reads
\beq
\label{bpw2}
P_{gb}(t,f') = \frac{dn(t,f')}{df'} {\pgw}_i\approx \frac{\Gamma_s\kappa^2}{4C_2G}t^{-3/2}f'^{1/2}.
\eeq
Further, proceeding in the way it is done to derive (\ref{rgw3}), we arrive to
\beq
\label{brgw3}
\rho_{bg} (t_0,f) = \frac{2\sqrt{2}}{4C_2G}\kappa^2\Gamma_sf^{1/2}H_r^{1/2}\int_{0}^{\infty}\frac{dz}{(1+z)^{5/2}}=
\frac{\sqrt{2}}{3C_2G}\kappa^2\Gamma_s\Omega_r^{1/4}H_0^{1/2}f^{1/2}.
\eeq
Finally, one can express the fraction (\ref{ogw1}) of the stochastic GWs background left over
by the domain walls escaping into wormholes as follows
\beq
\label{bogw2}
\Omega_{gb} (\ln f) = \frac{8\pi\sqrt{2}}{9C_2}\frac{\Omega_r^{1/4}}{H_0^{3/2}}\kappa^2\Gamma_sf^{3/2}.
\eeq

\section{Connecting to the NANOGrav signal}
\label{nanoAxion}

The NANOGrav~\cite{NANOGrav1} is sensitive to the characteristic strain (\ref{h1}) of GWs background
presented in terms of power low spectrum
\beq
\label{hNANO1}
h_c(f)=A_{\rm B}^{\rm yr}\left(\frac{f}{f_{\rm yr}}\right)^{\alpha},
\eeq
where $f_{\rm yr}=1{\rm yr}^{-1}=31$~nHz, $A_{\rm B}^{\rm yr}$ is the amplitude at $f_{\rm yr}$.
The spectrum (\ref{hNANO1}) is obtained in direct measurements of the timing-residual cross-power
density, whose slope is parametrized as $\gamma = 3-2\alpha$~\cite{NANOGrav1}. The fit (\ref{hNANO1})
is performed to thirty bins within frequency range from 2.5~nHz to 90~nHz. However, the
excess is reported in first five signal dominated bins spanning the range from 2.5~nHz to 12~nHz,
while the higher frequency bins are assumed to be white noise dominated. The signal excess
of the strain (\ref{hNANO1}) is reported for the parameters range
\beq
\label{A1}
-15.8\le\log A_{\rm B}^{\rm yr}\le -15.0
\eeq
 and
 \beq
 \label{G1}
 4.5\le\gamma\le 6.5
 \eeq
at 68\% confidence level (C.L.).

Using the formulas (\ref{psd1}), (\ref{h1}), (\ref{a1}) and (\ref{ogw2}) we calculate the signal strain of
GWs generated by the ALP field induced collapsing domain walls
\beq
\label{h2}
h_{c{\rm w}}^2(f)=C_4\Gamma_s\left(\frac{G\sigma }{f_{\rm yr}}\right)^2\left(\frac{f}{f_{\rm yr}}\right)^{-\frac{5}{2}},
\eeq
where
\beq
\label{C4}
C_4=\frac{16\sqrt{2}H_0^{1/2}\Omega_r^{1/4}}{7\pi C_2f_{\rm yr}^{1/2}} = 0.516.
\eeq
Therefore, comparing
\beq
\label{h3}
h_{c{\rm w}}(f)=0.72\sqrt{\Gamma_s}\left(\frac{G\sigma }{f_{\rm yr}}\right)\left(\frac{f}{f_{\rm yr}}\right)^{-\frac{5}{4}}
\eeq
with (\ref{hNANO1}) we find that slope parameter of the spectrum of stochastic GWs background generated in course of
the evolution of collapsing domain walls, induced by ALP inflationary dynamics, corresponds to
$\gamma = 5.5$, which exhibits a remarkable agreement with the range of values (\ref{G1}) reported in
NANOGrav signal~\cite{NANOGrav1}.

The stress energy density (\ref{sigma1}) gives rise the estimate
\beq
\label{Lambda1}
\Lambda = 0.6\Gamma_s^{-1/4}\sqrt{A_{\rm B}^{\rm yr}}M_{\rm Pl}\left(\frac{f_{\rm yr}}{F}\right)^{1/2},
\eeq
so that (\ref{A1}) implies~\footnote{$f_{\rm yr}=1.3\times 10^{-31}$GeV.}
\beq
\label{Lambda2}
\Lambda\approx 2.64\times 10^{-11}\Gamma_s^{-1/4}\left(\frac{C_A}{\rfh}\right)^{1/2}
\left(\frac{10^{13}{\rm GeV}}{\hinf}\right)^{1/2}{\rm GeV},
\eeq
where we put $A_{\rm B}^{\rm yr}=C_A\times 10^{-15}$ so that $C_A=0.16\div 1$.
Thus, the axion mass value needed to saturates the amplitude (\ref{A1}) of the NANOGrav signal lies at the level of
\beq
\label{mass1}
m_{\theta}^{\rm A}\approx 4.7\times 10^{-38}\Gamma_s^{-1/2}\left(\frac{C_A}{\rfh^2}\right)\left(\frac{M_{\rm Pl}}{\hinf}\right)^2{\rm eV},
\eeq
where $M_{\rm Pl}$ is taken for the Plank mass. Formula (\ref{mass1}) can be presented as
\beq
\label{mass12}
m_{\theta}^{\rm A}\approx 2\times 10^{-27}\frac{\Gamma_s^{-1/2}C_A}{\rfh^2}\ {\rm eV},
\eeq
where, to evaluate the numerical pre-factor, we applied the upper bound
\beq
\label{hinfl1}
\hinf =6\times 10^{13}\ {\rm GeV},
\eeq
as inferred from Plank results~\cite{plankX}, while the ratio $\rfh$ is still kept as a variable
which controls $\Gamma_s$ during the inflationary dynamics of the ALP field, as we will see in Section~\ref{infALP}.

Using  (\ref{psd1}), (\ref{h1}), (\ref{a1}) along with (\ref{bogw2}) one can also estimate the characteristic strain
of the signal generated by the acoustic response of bulk radiation fluid on the
formation of wormholes by domain wall with dominated gravitational dynamics
\beq
\label{hb1}
h_{c{\rm b}}(f) = 0.55\kappa\sqrt{\Gamma_s}\left(\frac{f}{f_{\rm yr}}\right)^{-\frac{1}{4}}.
\eeq
Therefore, with respect to (\ref{hNANO1}), the slope parameter of the acoustic
response spectrum corresponds to $\gamma = 3.5$, which is more than
$1\sigma$ off from the central value of the range (\ref{G1}). Obviously,
the total spectrum, around $f_{\rm yr}$, can be comparably
contributed by both $h_{c{\rm w}}(f)$ (\ref{h3}) and $h_{c{\rm b}}(f)$ (\ref{hb1}),
only in the case of valid relation
\beq
\label{both1}
G\sigma\simeq f_{\rm yr},
\eeq
which is not the case for the ALPs parameters localized in Section \ref{isocurv}.

\section{Inflationary dynamics of the ALP field}
\label{infALP}
The validity of derivation in Section~\ref{stochGW} is defined by the ability of domain walls
of the width $\simeq m_{\theta}^{-1}$ to oscillate their biggest fragments at the moment of the entering of the walls
into respective horizon scales. This implies that at the moment when the phase
$\theta$ finds itself in the oscillating state, the wave length corresponded
to the characteristic frequency band of the NANOGrav signal should exceed the inverse mass
of the ALP. The phase oscillations tun on when
\beq
\label{osc1}
m_{\theta}\approx H(z_{\rm osc}),
\eeq
so that, as follows from (\ref{hz2}),
\beq
\label{oscz1}
(1+z_{\rm osc})\approx \left(\frac{m_{\theta}}{H_r}\right)^{1/2}.
\eeq
The condition above simply demands that the ALP mass $m_{\theta}$
should exceed the blue shifted
characteristic scale of the NANOGrav, expressed as $(1+z_{\rm osc})f_{\rm yr}$.
Therefore, the lower bound on the mass of the ALP reads
\beq
\label{osc2}
m_{\theta} > f_{\rm yr}\left(\frac{f_{\rm yr}}{H_0}\right)\Omega_r^{-1/2}\approx 2.5\times 10^{-10}\ {\rm eV}.
\eeq

We assume that the inflation begins in a single Hubble volume
containing a phase value $\theta_0 < \pi$ which gets a random kick
of magnitude $\delta\theta$, given by (\ref{step1}), from vacuum quantum fluctuation of Fourier modes
leaving the horizon. In other words, the process of the phase variation can be
interpreted as a one dimension Brownian motion (random walk) of step length $\delta\theta$ (\ref{step1})
per Hubble time~\cite{antrop1,infl0,axInf2}.
Let us consider, in such a setup, the inflationary dynamics of the phase to define the formation rate $\Gamma_s$.
For every value of initial separation $\Delta\theta_{\pi}=|\pi -\theta_0|$,
one can define the probability density $p(\Delta\theta_{\pi},N)$ that
a Brownian path of the fluctuating phase will cross $\pi$ first time at a given e-fold
$N$, in a given Hubble volume $\hinf^{-3}$.
In general formulation, the first passage probability density is calculated as one
that a Brownian path, starting in $x_0$ at $t=0$, would cross the origin for the first time
at time $t$. This first-passage probability derived using
random walk~\cite{brown1} is given by
\beq
\label{p0}
p(x_0,t)=\frac{x_0^2}{\sqrt{4\pi D}}\frac{e^{-\frac{x_0^2}{4Dt}}}{t^{3/2}},
\eeq
where $D$ is interpreted as diffusion coefficient expressing the randomness.

In the language of the inflationary dynamics of the phase $\theta$ with respect to e-folds, we may use
the simplest one dimensional Pearson walk, which implies D=1/2, and re-define the special degree of
freedom in (\ref{p0}) as the phase difference
measured in steps $\delta\theta$, given by (\ref{step1}), so that
\beq
\label{p02}
x_0:=\Delta\theta_{\pi}/\delta\theta\approx 2\pi\Delta\theta_{\pi}\rfh .
\eeq
Since the quantum fluctuations increment (decrement) the phase value by $\delta\theta$, after each e-fold,
one has to use the e-fold as the time variable, in (\ref{p0}), so that $t:=N$.
In this setup the expression (\ref{p0}) is converted into
\beq
\label{p1}
p(\Delta\theta_{\pi},N)=(2\pi)^{3/2}\Delta\theta_{\pi}^2\rfh^2\frac{e^{-2\pi^2\frac{\Delta\theta_{\pi}^2}{N}\rfh^2}}{N^{3/2}},
\eeq
which will be used bellow to account for the probability of the domain walls' seeding contours formation
in the number density distribution (\ref{n1}).

As the first-passage happened, after $N$ e-fold from the beginning of inflation,
the Hubble volume, where it took place, became to be filled with the phase value
in the vicinity of $\pi$, so that the separation of the phase value
from $\pi$ should obey $\Delta\theta_{\rm first}\le\delta\theta$.
In this position of the phase, the probability density of $\pi$ crossing, in a Hubble
volume, at each e-fold, should be calculated from~(\ref{p1}) as
\beq
\label{p2}
p(\delta\theta ,1)\approx (2\pi e)^{-1/2}\approx 0.25.
\eeq
Therefore, the probability density $\Gamma_s$, corresponding to the horizon exit scale e-fold $N_s$,
can be factorized as
\beq
\label{Gsfact1}
\Gamma_s\approx p(\Delta\theta_{\pi},N_s-1)p(\delta\theta ,1).
\eeq
Obviously, that for all smaller scales, which correspond to later horizon exit e-folds $N_l>N_s$, $\Gamma_s$
can be taken as a constant, provided that $p(\Delta\theta_{\pi},N_s-1)<<1$.

The NANOGrav characteristic scale $f_{\rm yr}^{-1}$ exits the Hubble horizon at
\beq
\label{nyr1}
N_{\rm yr}=\ln (t_{\rm eq}f_{\rm yr}(1+z_{eq}))\approx\ln (t_{\rm eq}f_{\rm yr}) + \ln\frac{\Omega_m}{\Omega_r}
\eeq
e-fold, after the beginning of inflation. Using $t_{\rm eq}=51.1$~kyr~\cite{pdg1}~\footnote{The equality scale exits the inflationary
Hubble horizon at $N_{\rm eq}\approx\ln\frac{\Omega_m}{\Omega_r}= 7.96$.}
one arrives to the estimate $N_{\rm yr}\approx 18.8$.

Based on the above consideration, one can express the ALP's mass value (\ref{mass12}), saturating the
amplitude of NANOGrav signal, as a function of the initial phase separation
$\Delta\theta_{\pi}$ and the scale of the PQ symmetry breaking $\rfh$ measured
in units of the inflationary Hubble rate
\beq
\label{mass2}
m_{\theta}^{\rm A}(\Delta\theta_{\pi},\rfh ) = 6.0\times 10^{-28}C_A
\frac{(N_{\rm yr}-1)^{3/4}}{\Delta\theta_{\pi}\rfh^{3}}e^{\pi^2\frac{\Delta\theta_{\pi}^2}{N_{\rm yr}-1}\rfh^2}\ {\rm eV}.
\eeq

\begin{figure}
\includegraphics[width=0.5\textwidth]{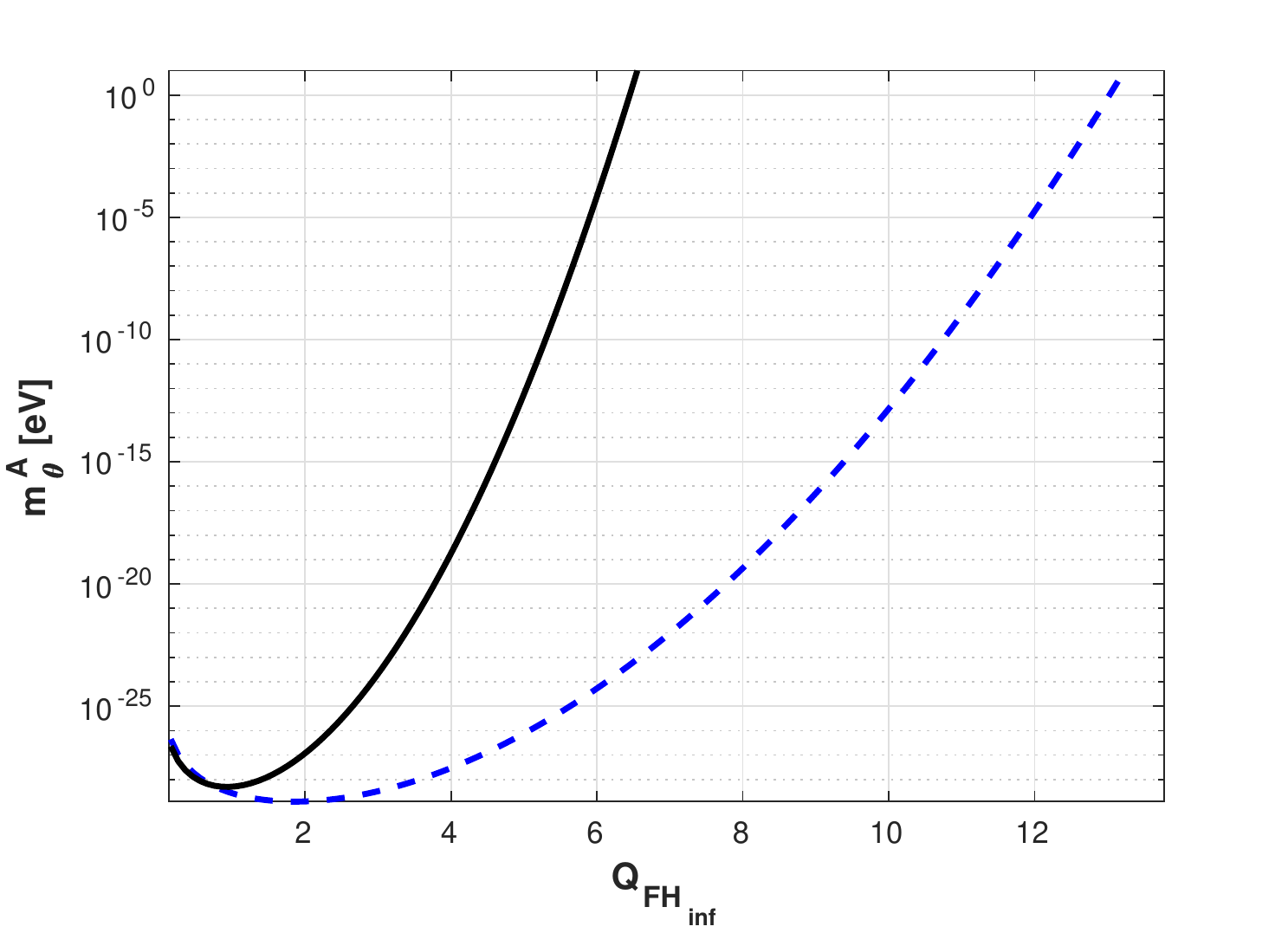}\hspace{0cm}\includegraphics[width=0.5\textwidth]{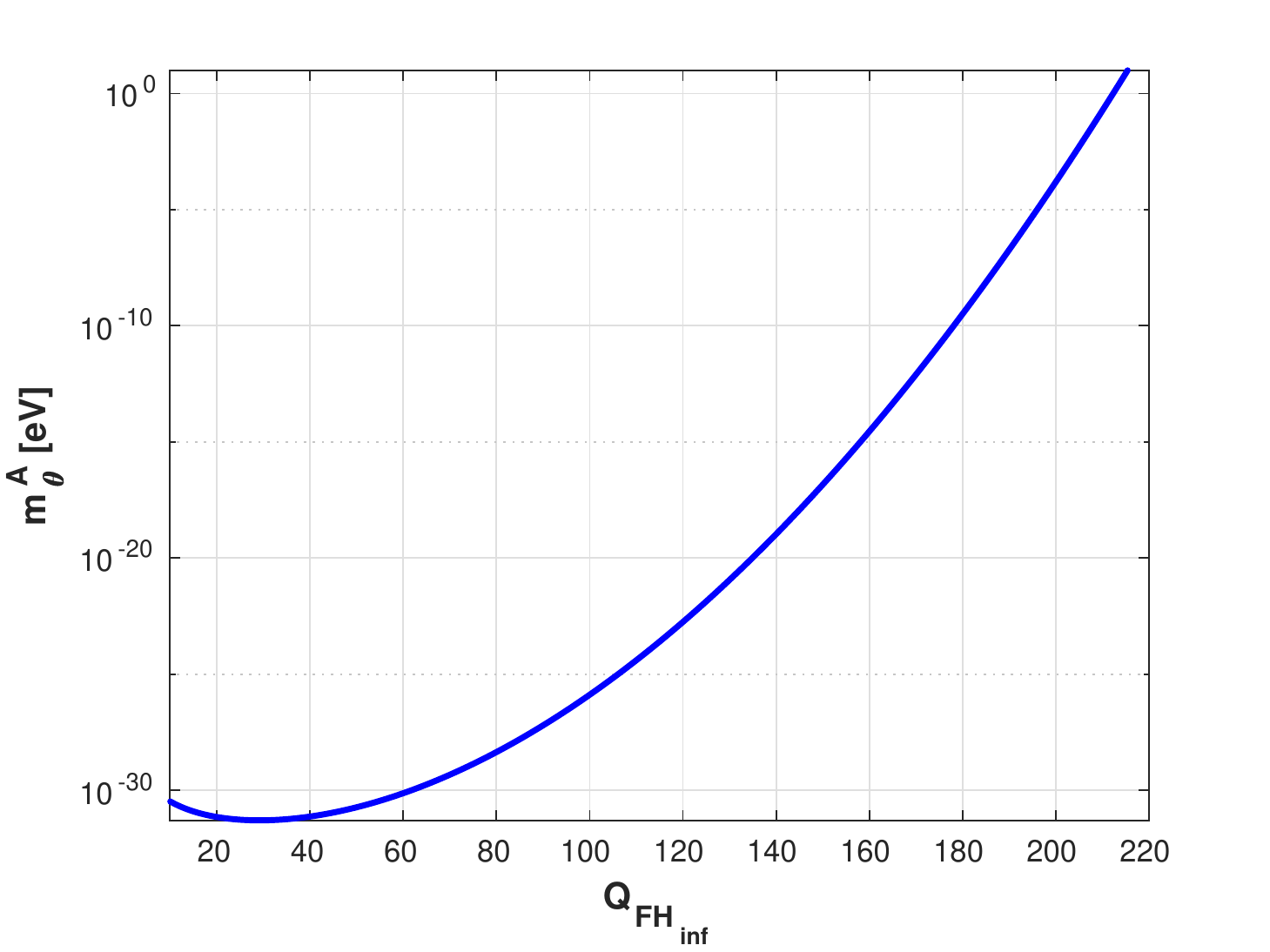}
\caption{\it The ALP's mass value~(\ref{mass2}) saturating the amplitude of the NANOGrav signal as a function of
the ratio $\rfh$. Left panel: the solid line corresponds to the initial phase value
$\theta_0\approx 10^{-2}$ chosen in the vicinity of the origin ($\Delta\theta_{\pi}=\pi$),
the dashed line corresponds to $\theta_0\approx \pi /2$ ($\Delta\theta_{\pi}=\pi /2$).
Right panel: the initial phase value $\theta_0$ is taken in the vicinity of $\pi$ ($\Delta\theta_{\pi}=0.1$).}
\label{fig:mA}
\end{figure}

The expression (\ref{mass2}) is visualized in Fig.\ref{fig:mA}
as the function of $\rfh$, for $C_A=1$ and three distinct values of the phase separation,
namely $\Delta\theta_{\pi}=\pi /2$ (left panel, dashed line), $\Delta\theta_{\pi}=\pi$
(left panel, solid line) and $\Delta\theta_{\pi}=0.1$ (right panel).

\section{Which ALPs may show up in the NANOGrav signal}
\label{isocurv}

Let us look at the ALP's inflationary dynamics related to the NANOGrav signal, as
described above, from the point of view of its simultaneous compatibility with constraints on the
DM density and isocurvature fluctuations.

In the biggest scale, emerged at the beginning of the inflation, from a single
causally connected domain of size $\hinf^{-1}$, the ALP field
is uniform and frozen at the initial phase value $\theta_0$, called misalignment
angle. The ALP remains frozen until the equality (\ref{osc1}) gets satisfied,
which takes place when the temperature of the Universe drops bellow certain
value $T_{\rm osc}$. In general, $m_{\theta}$ should depend on the temperature,
which, for example, in the case of the QCD
axion, is defined by the relation between $T_{\rm osc}$ and $\Lambda$ at which
the instanton effects become important (see the most advance lattice calculations
in~\cite{axionMass1}). For ALPs, such kind
of dependence should be driven by the details of its gauge couplings and
cannot be obtained though a straightforward generalization of the axion case.
Bellow, when we quantify the cosmological abundance of ALPs, we treat the mass
as being independent of temperature.
Our analysis is not sensitive to this assumption.

The total density of ALPs produced in the regime of the inflationary
spectator ALP field, namely under the condition (\ref{PQF}), is observationally
constrained by the CDM density. In particular, in the regime (\ref{PQF}), it is
required that the mass of ALP and its misalignment angle, defined by
$\theta_0$, would be mutually fine-tuned. Such kind of fine-tinning can be substantiated
with invoking of the anthropic reasoning, which initially has
been applied to the QCD axion~\cite{antrop1,antrop2}.
The anthropic selection, applied to the QCD axion~\cite{antrop1,antrop2},
implies that the misalignment angle should be small
($\theta_0<<1$) for PQ symmetry breaking scale relevant
for the inflationary QCD axion spectator.
In the version of the anthropic selection demanding that any
selected variables should have values near the maximal consistent with
existence of the human being, the CDM should completely consist of the QCD axions,
so that the anthropic window of parameter space is well defined for the QCD
axion~\cite{antrop1,antrop2}. However, in this anthropic window a large
isocurvature mode is strongly preferred~\cite{isoALPcurv1,antrop3,antrop4,isoALPcurv4}, which is
at odds with observations, as we discuss bellow. The above constraints
apply equaly to QCD  axion and ALPs. In particular, a multitude of ALPs
is expected to be produced in string theory framework attempting
to realize the axion scenario~\cite{alpString1,alpString2, alp2}, where
all ALPs can contribute into CDM content and produce
isocurvature perturbations. In this context, it is instructive to investigate, which ALPs
from the multitude, possibly existing in string theory, could be
seen as the source of stochastic GWs background being able to satisfy
the results of ALPs related NANOGrav signal compilation presented in Section~\ref{nanoAxion} and
Section~\ref{infALP}.

In Section~\ref{nanoAxion} we established that the slope parameter of the spectrum of
stochastic GWs generated by the domain walls,
created in the ALP inflationary dynamics,
exhibits a remarkably good agreement with that one reported by the NANOGrav.
Further, we specify the values of the ALP's scale $F$, its mass $m_{\theta}$ and
the misalignment angle $\theta_0$ which would lead to the saturation
of the NANOGrav signal amplitude (\ref{A1}) simultaneously satisfying the anthropic and
isocurvature constrains. In our study, we follow the QCD axion cosmology to describe the
origins of the constraints and quote the relevant formulas.
Except for the differences due to the treatment of the temperature-dependent mass, the formulas apply
equally well for the QCD axion and for ALPs.

When the ALP field starts to oscillate, the ALPs are produced
via vacuum misalignment mechanism~\cite{miss1,miss2,miss3} in form of cold Bose-Einstein
condensate with local number density defined by the initial phase value $\theta_0$
\beq
\label{axionDens1}
n_a\simeq\frac{1}{2}m_{\theta}F^2\theta_0^2f(\theta^2),
\eeq
where $f(\theta^2)\lesssim 10$ is a correction for anharmonic effects of the axion-like potential.
The number density $n_a$ red shifts in the same manner as the entropy density,
so that
\beq
\label{axionDens2}
\rho_a\simeq m_{\theta}s_0R_{\gamma}\frac{n_a}{s},
\eeq
where $s = (2\pi^2/45)g_{*s}T_{\rm osc}^3$ is the thermal entropy density when
the oscillations begin, while $s_0 = (2\pi^2/45)g_{*s0}T_0^3$ is the
nowadays entropy density $g_{*s0}= 3.91$ and $T_0=2.73$K.
The factor $R_{\gamma}$ indicates a possible entropy
production after the axion begins to oscillate.
Eventually, the above outlined misalignment ALPs abundance can be expressed as
(see for example~\cite{axionUnif1,axionUnif2})
\beq
\label{axionDens3}
\Omega_{\rm ALP}h^2\approx 0.12\left(\frac{5\times 10^{-9}\ {\rm eV}}{m_{\theta}}\right)^{1.165}
\left(\frac{\theta_0}{1.6\times 10^{-2}}\right)^2.
\eeq
This implies, that in the regime when the PQ-like symmetry remains broken during
the inflation and afterwards, ALPs in neV mass range will contribute
100\% into CDM, provided that the initial phase value obeys $\theta_0\lesssim 0.01$.
That way, the values of $m_{\theta}$ and $\theta_0$ define the
anthropic window for the QCD axion and ALPs.

The quantum fluctuations (\ref{step1}) do not alter the local energy density of the
ALPs, but instead are imprinted in the fluctuations of the ALPs number density
$\delta (n_a/s)\ne 0$, called isocurvature fluctuations.
Since the ALPs, considered here, are coupled
very weakly to the Standard Model (SM) particles~\footnote{The interaction to the
SM species is suppressed by factor $\propto F^{-1}$.},
they never come to the thermal equilibrium. In the absence of thermalisation,
these fluctuations must be compensated by radiation fluctuations. The amount
of ALP-like isocurvature perturbations $\alpha$ is constrained to
\beq
\label{iso1}
\alpha =\frac{\langle (\delta T/T)_{\rm iso}^2\rangle}{\langle (\delta T/T)_{\rm tot}^2\rangle}\lesssim 0.038
\eeq
at $k=0.05\ {\rm Mpc^{-1}}$, as inferred in~\cite{plankX}.
The details of estimation technique of the rate (\ref{iso1}), for QCD axion and ALP, are
given in~\cite{isoALPcurv1}. In particular, the relative contribution (\ref{iso1})
can be parametrized  as follows
\beq
\label{iso2}
\alpha =\frac{R_{\rm ALP}^2\langle S_{\rm ALP}^2(k)\rangle}{R_{\rm ALP}^2\langle S_{\rm ALP}^2(k)\rangle + \langle {\cal R}^2(k)\rangle},
\eeq
where $R_{\rm ALP}=\Omega_{\rm ALP}/\Omega_{\rm CDM}$, is the ALPs' fraction of
the total CDM content of the Universe, and $\langle {\cal R}^2(k)\rangle$ and $\langle S_{\rm ALP}^2(k)\rangle$
are the adiabatic and the entropy power spectra, respectively. The curvature power spectrum $\langle {\cal R}^2(k)\rangle$,
for the adiabatic mode of fluctuations of inflaton reads
\beq
\label{iso3}
\langle {\cal R}^2(k)\rangle =\frac{2\pi H_k^2}{k^3M_{\rm Pl}^2\epsilon_k},
\eeq
where $\epsilon$ is the first inflationary slow-roll parameter~\cite{infl1}
and the subscript indicates that the quantities are evaluated at $k=a\hinf$.
Provided, that the quantum fluctuations of the phase (\ref{step1}), are imprinted
with a nearly scale invariant spectrum
\beq
\label{iso4}
\langle\delta\theta^2(k)\rangle = \frac{2\pi^2}{k^3}\delta\theta^2
\eeq
one obtains~\cite{isoALPcurv1}
\beq
\label{iso5}
\langle S_{\rm ALP}^2(k)\rangle =\left\langle\left(\frac{\delta n_{\rm ALP}}{n_{\rm ALP}}\right)^2\right\rangle =
4\left\langle\left(\frac{\delta\theta}{\theta}\right)^2\right\rangle =\frac{2}{k^3\theta_0^2}\rfh^{-2}.
\eeq
Therefore, the isocurvature contribution (\ref{iso2}) can be related to the fundamental inflationary and
ALP's parameters by
\beq
\label{iso6}
\alpha\simeq\frac{R_{\rm ALP}^2\epsilon_k}{\pi\theta_0^2}\left(\frac{M_{\rm Pl}}{\hinf}\right)^2\rfh^{-2},
\eeq
where it is assumed that $\alpha\ll 1$, as follows from (\ref{iso1}).

\begin{figure}
\centering
\includegraphics[width=0.9\textwidth]{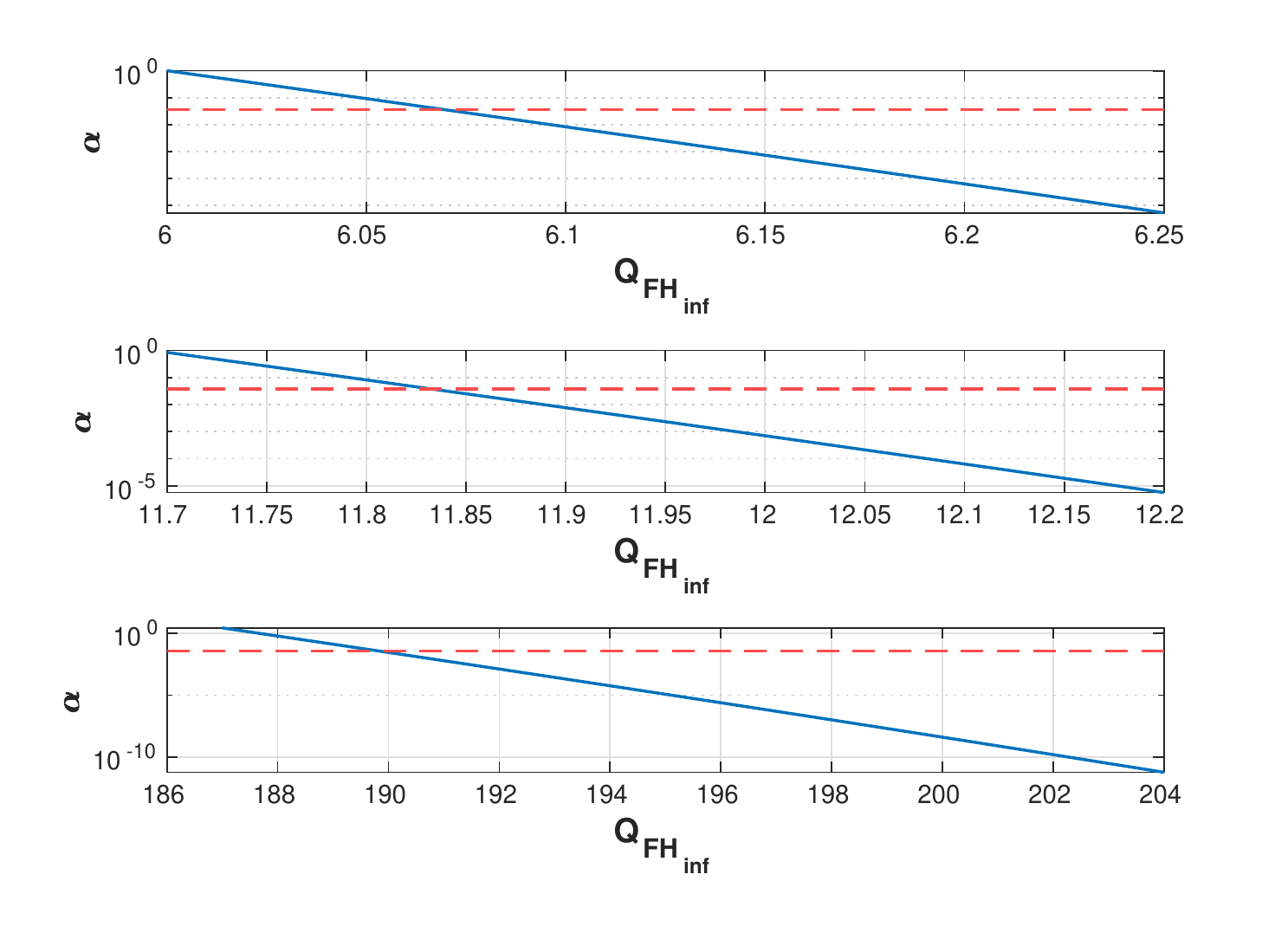}
\vspace{-0.4cm}
\caption{\it The fraction of ALP-type isocurvature perturbations as the function
of $\rfh$ (solid line) evaluated under the condition that the ALP mass saturates the
NANOGrav signal, as shown in Fig.\ref{fig:mA}. The upper panel corresponds to the initial phase value
$\theta_0\approx 10^{-2}$ ($\delta\theta_{\pi}\approx\pi$). The middle panel shows the functional dependence for
$\theta_0=\pi /2$ ($\delta\theta_{\pi}=\pi /2$). The lower panel corresponds to the initial
phase value in the vicinity of $\pi$, namely $\theta_0=\pi - 0.1$ ($\delta\theta_{\pi}=0.1$).
The horizontal, dashed line indicates the bound~(\ref{iso1}), deduced
from the Plank measurements.
}
\label{fig:iso}
\end{figure}

Taking into account (\ref{axionDens3}), one can express the isocurvature fraction (\ref{iso6}) as follows
\beq
\label{iso7}
\alpha (m_{\theta},\rfh ,\theta_0)\approx\left(\frac{5\times 10^{-9}\ {\rm eV}}{m_{\theta}}\right)^{2.33}
\left(\frac{\theta_0}{1.6\times 10^{-2}}\right)^4
\left(\frac{M_{\rm Pl}}{\hinf}\right)^2
\frac{\epsilon_k}{\pi\theta_0^2\rfh^2}
\eeq
During inflation, the primordial tensor perturbations are generated, setting
the initial amplitude for GWs which oscillate
after horizon entry. The spectrum  of these tensor perturbations is conveniently
specified as by the tensor fraction $r=\langle {\cal R}_T^2(k)\rangle/\langle {\cal R}^2(k)\rangle$.
In the slow-roll approximation~\cite{infl1}
\beq
\label{r1}
r=16\epsilon
\eeq
is experimentally bounded to $r<0.058$~\cite{pdg1} (at $k=0.002\ {\rm Mpc^{-1}}$), which we use to estimate the upper
limit on the first slow-roll parameter
\beq
\label{r2}
\epsilon_k\approx 0.036.
\eeq

Evaluating (\ref{iso7}), under the condition of saturation of the NANOGrav amplitude (\ref{A1}),
which implies that $m_{\theta}=m_{\theta}^A$ (\ref{mass2}),
and for the upper bounds (\ref{hinfl1}) and (\ref{r2}), one can frame three scenario in which
the ALP might contribute, by its inflationary dynamical induced formation of the domain walls,
into the stochastic GWs indicated in the NANOGrav observations.

{\bf (i)} In the first scenario, the initial value of the phase, during inflation,
might be chosen in the vicinity of the origin $\theta_0\approx 10^{-2}$
($\Delta\theta_{\pi}\approx\pi$), which would correspond to the anthropic
window usually discussed in the context of the QCD axion. The bound (\ref{iso1}), as
it is displayed in Fig.\ref{fig:iso} (upper panel),
allows to accommodate ALPs with their inflationary dynamics driven by the ratio
\beq
\label{alpQ1}
\rfh\ge 6.1
\eeq
which corresponds~\footnote{Provided that typical amplitude of quantum fluctuations (\ref{step1}),
at given value of ratio (\ref{alpQ1}), is about $10^{-2}$, the initial position of phase $\theta_0$,
in this scenario, is not significantly disturbed, at the biggest scales.},
at bound (\ref{hinfl1}), to
\beq
\label{alpF1}
F\ge 3.8\times 10^{14}\ {\rm GeV.}
\eeq
In this scenario, according to (\ref{mass2}), as it is shown in left
panel of Fig.\ref{fig:mA}, ALP with detectable domain walls GWs signature should have mass
\beq
\label{alpMass1}
m_{\theta}\gtrsim 10^{-4}\ {\rm eV}
\eeq
and contribute, as it follows from (\ref{axionDens3}),
\beq
\label{alpFRAC1}
R_{\rm ALP}\lesssim 10^{-5}
\eeq
of the total CDM content of the Universe. Eventually, applying (\ref{sigma1}), one obtains
the value of the stress energy density of the domain walls
\beq
\label{sigma0}
\sigma\gtrsim 5.8\times 10^{16}\ {\rm GeV^3}.
\eeq

{\bf (ii)} The second scenario assumes that the initial phase value
$\theta_0=\pi /2$ (that is $\mathcal{O}(1)$), and, as it is shown in Fig.\ref{fig:iso} (middle panel),
accommodates the ALP with ratio
\beq
\label{alpQ2}
\rfh\ge 11.9,
\eeq
which corresponds to
\beq
\label{alpF2}
F\ge 7.0\times 10^{14}\ {\rm GeV.}
\eeq
In this scenario, the ALP with detectable domain walls implies
\beq
\label{alpMass2}
m_{\theta}\gtrsim 1.2\times 10^{-5}\ {\rm eV},
\eeq
as it follows from (\ref{mass2}) (see left
panel of Fig.\ref{fig:mA}).
Thus, according to  (\ref{axionDens3})
\beq
\label{alpFRAC2}
R_{\rm ALP}\lesssim 1,
\eeq
which implies that the ALPs constitutes total CDM budget of the
Universe. The value of the stress energy of the domain walls reads
\beq
\label{sigmapi2}
\sigma\gtrsim 2.3\times 10^{16}\ {\rm GeV^3}.
\eeq

{\bf (iii)} In the third scenarios, we put the initial phase value to $\theta_0=\pi - 0.1$ ($\Delta\theta_{\pi}=0.1$),
which implies the initial phase position $\theta_0$ in a vicinity of the domain wall formation crossing point.
Thus, the isocurvature bound (\ref{iso1}), as shown in lower panel of Fig.\ref{fig:iso}, can
accommodate the ALP with ratio
\beq
\label{alpQ3over}
\rfh\ge 189.8,
\eeq
which would correspond to the ALP mass, saturating the NANOGrav signal amplitude
(see Fig.\ref{fig:mA}, right panel),
\beq
\label{alpMAss3over}
m_{\theta}\gtrsim 1.7\times 10^{-7}\ {\rm eV}.
\eeq
However, according to (\ref{axionDens3}), the ALPs with such mass and initial
phase, in the vicinity of $\pi$, are definitely overabundant.
If we increase the mass limit up to
\beq
\label{alpMAss3}
m_{\theta}\gtrsim 4\times 10^{-5}\ {\rm eV},
\eeq
which implies that (see Fig.\ref{fig:mA}, right panel)
\beq
\label{alpQ3}
\rfh\ge 198.2
\eeq
and
\beq
\label{alpF3}
F\ge 1.2\times 10^{16}\ {\rm GeV},
\eeq
the ALPs will constitute the total CDM content of the Universe ($R_{\rm ALP}\lesssim 1$)
and give a negligible contribution into the ALP-like
isocurvature fraction, which reads
\beq
\label{alpha3}
\alpha\lesssim 7\times 10^{-8}.
\eeq
In this case, one expects the highest possible value of the stress
energy density of the domain walls, namely
\beq
\label{sigmapi}
\sigma\gtrsim 5.7\times 10^{18}\ {\rm GeV^3}.
\eeq

Thus, in the context of detectability of ALPs from their multitude, possibly produced
in string theory~\cite{alpString1,alpString2, alp2} along with the QCD axion,
one infers the following.  If one
of the ALPs has the misalignment angle value set to the same
one, $\theta_0\approx 10^{-2}$, imposed by the anthropic window of the QCD axion,
it may saturate the NANOGrav excess amplitude even having a negligible
contribution into the CDM density (\ref{alpFRAC1}), at the same time being consistent with the
isocurvature constraint. If one of the ALPs has large value of the
misalignment angle, $\theta_0\approx 1$, it should dominate the CDM density (\ref{alpFRAC2}) and the
isocurvature contributions to be interpreted as the source of the saturation of the NANOGrav amplitude.
In case of one of the ALPs has $\theta_0$ value very close to $\pi$, which is, however,
as probable as the QCD axion anthropically selected $\theta_0$, it may manifest itself in the NANOGrav,
in the condition of the dominant contribution into the CDM density and negligible
contribution into the isocurvature perturbations (\ref{alpha3}). However, in this case,
instead of collapsing the domain walls of NANOGrav scale range should form wormholes,
as we elucidate in Section~\ref{sec:bu}.

Let us notice that, according to (\ref{passN2}), at
$N\approx 60$, the domain wall formation crossing point is reached within
almost 100\% of the horizon exiting scales, for each of three above considered scenario.
However, these scales, corresponding to the very end of inflation, are definitely
much smaller than the width of domain walls typical for the scenario described above,
so that the dynamics of the resulting vacuum-like objects should be much different
than that one of the domain walls discussed above.


\section{Domain walls escaping into baby universes}
\label{sec:bu}

Walls of size $R(t_H)=H^{-1}$ cross the Hubble radius defined by the Hubble rate $H=\dot a/a$.
The mass of a wall at Hubble crossing is expressed as
\beq
\label{m1}
M_{\rm w}(H)\approx 4\pi\sigma H^{-2}=4\pi M_{\rm Pl}\left(\frac{\sigma}{M_{\rm Pl}^3}\right)
\left(\frac{M_{\rm Pl}}{H}\right)^2.
\eeq
Once the wall becomes encompassed by a Hubble radius it collapses, within
about a Hubble time, into a PBH of mass $M_{\rm PBH}=\xi M_{\rm w}$, where $\xi\simeq 1$
indicates the fraction of the wall energy deposited into the BH.
The wall stress energy tension is a source of repulsive gravity~\cite{wall1,wallEvol1,wallEvol2,wallGW0} and hence
should maintain a negative pressure in case of a domination of the wall's
material inside of encompassing it Hubble radius.
The collapse can take place unless the the mass of the wall inside a
given Hubble radius exceed the mass of its matter content which happens
when the Hubble rate reaches the value
\beq
\label{Hw1}
H_{\rm w}\approx 8\pi\sigma G = 8\pi M_{\rm Pl}\left(\frac{\sigma}{M_{\rm Pl}^3}\right).
\eeq

Thus, for stress energy density (\ref{sigmapi}) relevant for the ALP parameters localized in scenario (iii), discussed
in Section~\ref{isocurv}, the dominating wall ``enters''~\footnote{It is more likely to say, that the size
of the wall coincides with the Hubble radius.}
 the horizon at Hubble rate $H_{\rm w(iii)}\approx 10^{-9}\ {\rm eV}$.
Comparing this rate with that one at which $f_{\rm yr}$ enters into Hubble radius (\ref{osc2}) one
can see that the walls of size corresponding to the frequencies $f_{\rm w(iii)}\gtrsim 300$~nHz are able
to collapse into PBHs.  Bigger walls entered the Hubble radius starts expanding faster than the background reaching
eventually the inflationary vacuum and develop wormholes to baby universes~\cite{wallPBH1,wallPBH2}.
Such wormholes are seen as BHs in the FRW Universe.
Since this frequency is much higher than the frequencies of the
signal bins of the NANOGrav, the ALP domain walls, described in (iii), should mostly
form wormholes. Provided that the condition (\ref{both1}) fails in each
scenario of Section~\ref{isocurv}, the acoustic response signal (\ref{hb1}) induced
by these escaping walls cannot saturate the amplitude (\ref{a1}).
Therefore, most likely the ALP field with parameters
specified in scenario (iii) cannot be a source of the signal indicated by the NANOGrav.

In scenarios (i) and (ii), the dominating wall enters the horizon at
Hubble rate $H_{\rm w(i)}\approx 10^{-11}\ {\rm eV}$, which corresponds
to the GW signal frequency of $f_{\rm w(i)}\lesssim 3$~nHz. Therefore,
the domain walls induced by inflationary dynamics of the
ALP fields specified in scenarios (i) and (ii) should
collapse into PBHs within almost the whole range of frequency support
of the NANOGrav. The slope of the strain spectrum (\ref{h3}), which might be
generated by these walls, $\gamma =5.5$, exhibits a remarkable agreement
with the central value of the slope range (\ref{G1}) reported by NANOGrav~\cite{NANOGrav1},
in its excess frequency bins. Generically, it would be reasonable to expect a
sort of spectral feature expressed  as the slope change from $\gamma =3.5$,
for frequencies bellow $f_{\rm w(i)}\simeq 3$~nHz to $\gamma =5.5$ for
higher frequencies. Such kind of feature might indicate
the transition from the regime when bigger walls, ``entering''
their Hubble radii, were escaping into baby universes and
those smaller ones which were collapsing into PBHs.  However,
because of the amplitude inconsistency between collapsing (\ref{h3})
and escaping (\ref{hb1}) strains, caused by the violation
of condition (\ref{both1}),~\footnote{In all 3 scenarios of section \ref{isocurv}, $G\sigma >> f_{\rm yr}$.}
one might expect to see only a spectrum of slope $\gamma =5.5$ with
amplitude almost abruptly attenuating at frequencies
bellow $f_{\rm w(i)}\simeq 3$~nHz.

Currently, NANOGrav has about $T\approx 15$ years
of high precision timing observations from
many pulsars, where every pulsar is observed each $\Delta t\simeq 1\div 3$ weeks
with integration time about 20 minutes. Thus, the sensitivity band to GW frequency
can be defined as $1/T<f<\Delta t/2$, which spans the range from $\simeq 2$~nHz
to $\simeq 1\ {\rm\mu Hz}$. This means, that in scenarios (i) and (ii),
the frequency range indicating the domain walls forming wormholes, may be,
just barely reaches the lowest frequency bin accessible for the NANOGrav.
Therewith, it would be hard to expect to obtain a real testification
on wormholes formation by ALP field induced domain walls.

The boundary between a PBH collapsing and wormhole escaping domain wall
can be characterized  by the mass of wall material contained within the size $H_{\rm w}$,
which reads
\beq
\label{Hw2}
M_{\rm w}(H_{\rm w})\approx\frac{M_{\rm Pl}}{16\pi}\left(\frac{M_{\rm Pl}^3}{\sigma}\right).
\eeq
For the parameters specified for the NANOGrav detectable ALPs, the values of the boundary mass
are $M_{\rm w(i)}\approx 7{\rm M_{\odot}}$ in scenario (i),
$M_{\rm w(ii)}\approx 18{\rm M_{\odot}}$ in scenario (ii) and
$M_{\rm w(iii)}\approx 0.07{\rm M_{\odot}}$ in scenario (iii). These values
indicate the upper bounds of PBH mass formed by collapsing domain walls. The heavier
BHs are induced by domain walls developing wormholes, so that their masses
cannot exceed the energy of the radiation fluid contained in a respective
Hubble radius~\cite{wallPBH1}, provided that horizon entering took place before
the equality epoch.

Closing the section, we notice, that the observational access to the signature of domain walls
escaping into baby universes may be
more encouraging in the context of the spontaneous nucleation of domain walls on the inflationary
stage, considered in~\cite{vilenkin1,wallPBH1,wallPBH2}. In this setup, the nucleation
rate $\Gamma_s\propto\exp(-S_E)$ is defined by the action $S_E$ of the semi-classical
tunneling pass, $S_E\simeq 2\pi^2\sigma\hinf^{-3}$. To instance, for higher reference frequency
$f_0>f_{\rm yr}$, one may represent (\ref{hb1}) as
\beq
\label{hb2}
h_{c{\rm b}}(f) = 0.55\kappa\sqrt{\Gamma_s}\left(\frac{f_{\rm yr}}{f_0}\right)^{\frac{1}{4}}
\left(\frac{f}{f_{\rm yr}}\right)^{-\frac{1}{4}}.
\eeq
Therefore, the signature of domain wall escaping into baby universes
might be seen with a detector of GWs having sensitive to amplitude
\beq
\label{a2}
A_B^0\lesssim  0.3\kappa^2\Gamma_s\left(\frac{f_{\rm yr}}{f_0}\right)^{\frac{1}{2}}
\eeq
in the characteristic strain power spectrum like (\ref{hNANO1}), measuring the slope value
to be close to that one of the acoustic response, $\gamma =3.5$. Presumably,
Gaia and THEIA can have some potential to provide sensitive measurements in
much higher frequency band~\cite{detGW1}.

\section{Consistency of collapsing walls evolution with PBHs constraints}
\label{sec:bh}

There is no domain wall problem related to their production mechanism considered
in this paper. The walls which create wormholes export the domain wall problem
into baby universes. The collapsing walls form PBHs before their contribution
into the energy density becomes large enough to contradict the observational constraints.

The dynamics of spherical domain wall has been studied in~\cite{wallEvol1}.
The result of this study implies that a closed domain wall being initially bigger than its Schwartzshild radius always collapses
into a BH. Thus, the ALP field induced domain wall after obtaining of a spherical geometry contracts toward the
center. Since the ALP's coupling to the Standard Model particles is strongly suppressed
by the large scale $F$, so that the ALP wall interacts very weakly with matter and hence nothing can prevent
the wall finally to be localized within a volume of size comparable to its width~\cite{we1,we2}.
The Schwarzschild radius of a wall of total mass $M_{\rm w}$
\beq
\label{swR1}
R_{\rm S}=2GM_{\rm w}
\eeq
can be expressed trough the radius of the wall $R_{\rm w}$ and the ALP parameters as follows
\beq
\label{swR2}
R_{\rm S}=32\pi GR_{\rm w}^2m_{\theta}F^2.
\eeq
Provided that the wall contracts down to about
the size comparable to its width $\simeq m_{\theta}^{-1}$, one naturally assumes that
a PBH will be formed under the condition $R_{\rm S}>m_{\theta}^{-1}$, which reads
\beq
\label{swR3}
R_{\rm w}>\frac{m_{\theta}^{-1}}{4\sqrt{2\pi}}\frac{M_{\rm Pl}}{F}
\eeq
Since, the NANOGrav frequencies would correspond to the size of the emitting object
\beq
\label{swR4}
R_{\rm w}\simeq\frac{f_{\rm yr}^{-1}}{1+z_{\rm osc}},
\eeq
one can impose the following constraint on the ALP's mass
\beq
\label{swR5}
m_{\theta}\gtrsim\frac{32\pi}{\sqrt{\Omega_{\rm r}}}f_{\rm yr}
\left(\frac{f_{\rm yr}}{H_0}\right)\left(\frac{M_{\rm Pl}}{F}\right)^2
\approx 8.2\times 10^{-16}\left(\frac{M_{\rm Pl}}{F}\right)^2\ {\rm eV},
\eeq
where we used (\ref{oscz1}) to evolve (\ref{swR3}) and (\ref{swR4}).
The numerical value of the constraint (\ref{swR5}) reads
$m_{\theta {\rm (i)}}\approx 8.5\times 10^{-7}$~eV, $m_{\theta {\rm (ii)}}\approx 2.5\times 10^{-7}$~eV
and $m_{\theta {\rm (ii)}}\approx 8.5\times 10^{-10}$~eV for scenario (i), (ii) and
(iii) respectively. Thus, one can affirm that in all three scenarios described in Section~\ref{isocurv},  the condition (\ref{swR5})
is well satisfied, which implies that the domain walls producing GWs signals of frequencies
about $f_{\rm yr}$, are capable to deposit their energy under their Schwarzschild radius and hence
convert it into PBHs. It is obvious, that the  above statement is still in tact
also for frequencies one order of magnitude bellow and above $f_{\rm yr}$.

In what follows, we verify the consistency of PBHs abundance of
different masses produced by collapsing closed domain walls, with existing
constraints~\cite{Caretal20}.

\begin{figure}
\includegraphics[width=0.5\textwidth]{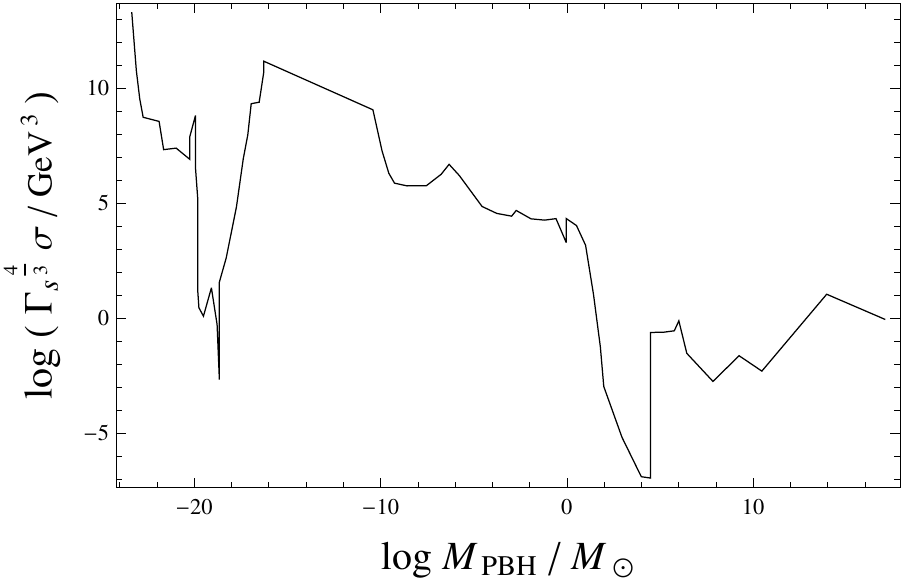}\hspace{0cm}\includegraphics[width=0.5\textwidth]{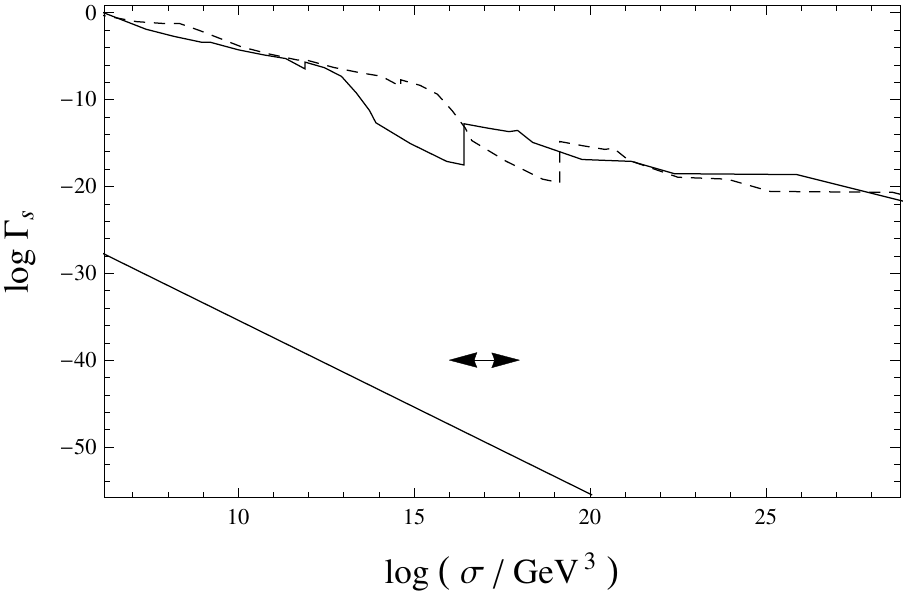}
\caption{\it Left panel: Constraints on the combination of parameters
$\sigma\Gamma_s^{4/3}$ based on the known constraints on primordial black holes~\cite{Caretal20}.
Right panel:  The solid curve corresponds
to the observed NANOGrav frequency $f_0=2.5$~nHz,
the dashed curve corresponds to $f_0=12$~nHz.
The straight line shows the NANOGrav driven correlation (\ref{GsTOsigma}) for $C_A=1$.
The double side arrow sign corresponds to the range of the domain walls stress energy defined in scenarios
(i)-(iii) of Section~\ref{isocurv}.}
\label{fig:bh}
\end{figure}

The mass distribution of the BHs is defined by the size distribution of the walls' seeding contours (\ref{n1})
scaled with respect to the expansion of the Universe, which is given in (\ref{n2}) and reads as
\begin{equation}
dn=\Gamma_s\frac{dR}{t_{eq}^{3/2}R^{5/2}},
\label{raspr}
\end{equation}
at the equality time. A useful characteristic of this distribution, which allows to
compare the PBHs yield with the constraints on their abundance in different mass
range~\cite{Caretal20}~\footnote{In particular, see Fig. 18 in~\cite{Caretal20}.},
is the mass density of PBHs per logarithmic mass interval
in units of the total density of the Universe
\beq
\label{raspr1}
\frac{d\Omega_{\rm PBH}}{d\ln M_{\rm PBH}}=\frac{1}{\rho_{eq}}\frac{dn}{d\ln M_{\rm PBH}}M_{\rm PBH},
\eeq
where $\rho_{\rm eq}=M_{\rm Pl}^2/(6\pi t_{\rm eq}^2)$ is the matter density at the time of equality.
Using (\ref{raspr}), we can obtain
\beq
\label{raspr2}
\frac{dn}{d\ln M_{\rm PBH}}=\frac{\Gamma_s}{2}\frac{1}{t_{eq}^{3/2}}(4\pi\sigma)^{3/4}M_{\rm PBH}^{-3/4},
\eeq
so that (\ref{raspr1}) can be converted into
\beq
\label{raspr3}
\frac{d\Omega_{\rm PBH}}{d\ln M_{\rm PBH}}=\frac{3\pi \Gamma_s t_{\rm eq}^{1/2}(4\pi\sigma)^{3/4}M_{\rm PBH}^{1/4}}{M_{\rm Pl}^2},
\eeq
to be compared with the model
\beq
\label{contrPBH1}
\Omega_{\rm PBH}\sim10^9\beta\left(\frac{M_{\rm PBH}}{M_\odot}\right)^{-1/2}
\eeq
used in~\cite{Caretal20} to quote the constraints on the density fraction $\beta$ deposited in PBHs at the moment of their formation.
In particular, independently of the NANOGrav signal, one can provide an upper bound, shown in the left panel of Fig.~\ref{fig:bh},
on the combination $\sigma\Gamma_s^{4/3}$ for PBHs of different masses formed
in a domain walls collapse process, provided that each wall converts its total mass into a PBH.
One can see that the restrictions (see left panel in Fig.~\ref{fig:bh}) become gap-like
tougher, namely $\sigma\Gamma_s^{4/3}\le 10^{-7}$, in a range of masses exceeding the boundary mass
$M_{\rm w}(H_{\rm w})=17M_{\odot}$ quoted above.

In general, the relation (\ref{m1}) implies that different combinations of $R$ and $\sigma$ may be composed into one
and the same mass of the PBH. This degeneracy is removed if we fix the reference frequency $f_{\rm yr}$ of the NANOGrav signal. Thus,
using the chain of relations (\ref{raspr3}), (\ref{m1}), $R=ct$,
\begin{equation}
1/t=f=f_{\rm yr}(1+z),
\end{equation}
and the connection
\begin{equation}
1+z=(1+z_{\rm eq})t_{\rm eq}^{1/2}/t^{1/2}
\end{equation}
one arrives to
\begin{equation}
\sigma=\frac{M_{\rm PBH}f_0^4t_{\rm eq}^2(1+z_{\rm eq})^4}{4\pi},
\label{seq}
\end{equation}
and
\begin{equation}
\Gamma_s=\frac{(d\Omega_{\rm PBH}/d\ln M_{\rm PBH})M_{\rm Pl}^2}{3\pi  t_{\rm eq}^2 M_{\rm PBH} f_0^3(1+z_{\rm eq})^3}.
\label{gsrestr}
\end{equation}
The expressions (\ref{seq}) and (\ref{gsrestr})
together with constraints presented in~\cite{Caretal20} impose the bound on $\Gamma_s$,
which is shown by dogleg lines in the right panel of Fig.~\ref{fig:bh}.

Using (\ref{h3}) and (\ref{A1}) one can infer the mutual correlation which
NANOGrav signal imposes on $\sigma$ and $\Gamma_s$, in case of saturation of the signal amplitude
\beq
\label{GsTOsigma}
\log\Gamma_s = 2\log C_{\rm A} -15.4 -2\log\left(\frac{\sigma}{1{\rm GeV}^3}\right)
\eeq
This equation is shown by the straight line in the right panel of Fig.~\ref{fig:bh}.

Thus, one can see that the PBHs contribution from the collapsing ALP domain walls
saturating with their GWs emission the NANOGrav signal is well bellow the allowed
limit on the PBHs abundance.

\section{Conclusions}
\label{sec:conclusions}

In this paper, we have explored the impact of the evolution of closed domain walls,
created in course of dynamics of ALP field spectating the inflation, on the
stochastic GWs background in the frequency range accessible
for the PTA measurements.

As long as the Universe is inflating, an axion or ALP field owing,
by definition, the discrete set of vacuum states,
with PQ~(-like) symmetry broken before inflation, experiences
the quantum fluctuations. In some patches of the space, an
accumulation of the fluctuations can
move the field, from its initial value (misalignment angle), so much
that after inflation it will be located in a vacuum state
differing from that one corresponding to its
value occurred at the beginning
of the inflation. Thus, due to inflationary expansion and
provided that the probability for the field to acquire
the value corresponding to another vacuum state is small enough,
one expects to see a Swiss cheese like picture where
the domains with underrepresented vacuum will be
inserted into a space remaining filled with the initial one.
Two different neighboring vacuum states should be interpolated
by a domain wall, represented by the sine-Gordon kink solution (\ref{sinG1})
in the case of axions or ALPs. Obviously, the Swiss cheese
structure guaranties that the produced domain walls
have closed, although irregular,  shapes~\footnote{There are non-axion related realizations of the
mechanism of closed domain walls formation relied on the
inflationary dynamics in a landscape, see for example~\cite{wallG1,wallG2}.}.

Evolving after inflation, the domain walls, depending on their sizes,
either collapse forming PBHs or escape into baby universes leaving
wormholes anyway observed as BHs in our Universe. Therethrough,
there is no domain wall problem in either scenario.

The collapsing walls tend to decrease their entropy leading to a smoothing out of
their surfaces, and hence radiating our the energy in form of
GWs, with characteristic frequency of about the Hubble rate
established during their collapsing instants. We have estimated the characteristic strain power spectrum
produced by the size distribution of the collapsing closed domain walls and
relate it with the recently reported NANOGrav signal excess obtained
in PTA measurements. It is remarkable, that the slope of the frequency dependence of the strain
spectrum $\gamma =5.5$, generated by such domain walls, is very well centered inside
the range of the slopes in the signal obtained by the NANOGrav.

Analyzing the inflationary dynamics of the ALP
field, in consistency with the isocurvature constraint, we defined those combinations of its
parameters where the signal from the inflationary induced ALPs domain walls could saturate
the amplitude of the NANOGrav excess. In particular, if an ALP
has the misalignment angle value set to the same one, imposed by the anthropic
window of the QCD axion, it may saturate the NANOGrav excess amplitude even having a
negligible contribution into the CDM density, at the same time being consistent with the
isocurvature constraint. Provided that  string theory models that include the QCD axion,
generically at high scale of PQ symmetry breaking, might also incorporate other ALPs
which can leave their imprint in the stochastic GWs background, with
the preferred by the NANOGrav excess parameters of the characteristic strain power spectrum and
without overproduction of CDM and/or isocuvature modes.

Those walls exceeding by their mass the energy of the bulk fluid inside the Hubble radius
will dominate by their repulsive nature and hence push the fluid away, leaving two
nearly empty layers in their vicinity. Such walls, keeping exponentially extending,
escape into baby universes forming wormholes in our Universe, which are observed as
BHs of masses ranging from planetary to some of LIGO/Virgo scales~\cite{ligo1,ligo2}, depending
on the parameters of ALPs. We assumed that being pushed away
the bulk radiation fluid responds
by acoustics waves which may set in a motion a fraction of the mass in Hubble radius.
This mass, in a motion, could generate the GWs signal with the strain spectrum slope
and the amplitude different from that ones expected from the collapsing walls.
It was found that in the context of domain walls produces by ALPs, it is quite unlikely,
while in principle possible, to trace baby universes in PTA measurements. However,
a detectable stochastic GWs imprint of the phenomena can be specified for other types
of phase transitions which might occur during inflation.

At some final inflationary e-folds, corresponding to much smaller than the NANOGrav scales, the probability of the ALP field to move its
value into another vacuum may become large, which may correspond to percolated
domain walls system considered in~\cite{wallSmall1} and induce small scale density inhomogeneities
in the distribution of the coherent oscillations of the ALPs Bose-Eisenstein condensate.
Possible scenarios of evolution and observational signatures of small scale
inhomogeneities in axionic Bose-Eisenstein condensate have been studied in great details in the cosmology of QCD axion
(see for example~\cite{becSmall1,becSmall2,becSmall3,becSmall4,becSmall5,becSmall6}).

\section*{Acknowledgments}
The work of Sergey G. Rubin has been supported by the Kazan Federal
University Strategic Academic Leadership Program,  by
the Ministry of Science and Higher Education of the Russian Federation, Project
``Fundamental properties of elementary particles and cosmology''
N~0723-2020-0041 and by RFBR grant N~19-02-00930.


\appendix
\section{First passage e-fold}
\label{sec:first}
Providing the condition $p(\Delta\theta_{\pi},N)\approx 1$, in (\ref{p1}),
one can define how many e-folds would it take the phase to cross $\pi$, solving the following
equation
\beq
\label{eqN1}
3N_{\pi}^2+(\frac{1}{2}\ln 2\pi+\ln \Delta\theta_{\pi} +\ln\rfh-\frac{3}{2}\ln N_{\pi})N_{\pi}-2\pi^2\Delta\theta_{\pi}^2\rfh^2=0.
\eeq
One can put the term $\frac{3}{2}\ln N$ in the parenthesis to a constant $C_1\approx 4.5\div 6.2$,
provided that it changes quite slow with increasing
of $N$ in the range $N\approx 20\div 60$.
Under this assumption, the solution of (\ref{eqN1}) can be approximated by the root of quadratic equation, given by
\beq
\label{passN1}
N_{\pi}\approx\frac{(24\pi^2\Delta\theta_{\pi}^2\rfh^2-(0.9\ln\Delta\theta_{\pi}+\ln\rfh -C_1))^{1/2}}{6}-
\frac{(0.9\ln\Delta\theta_{\pi}+\ln\rfh -C_1)^2}{6}.
\eeq
For reasonable choose of parameter, $1<\rfh\lesssim 10$ and $\Delta\theta_{\pi}\simeq \pi /2$ and provided that
typically accepted number of e-folds needed for inflation $N_{\rm inf}\simeq 60$, one can neglect relevant terms in
(\ref{passN1}), so that the solution is reduced to
\beq
\label{passN2}
N_{\pi}\approx\pi\Delta\theta_{\pi}\rfh .
\eeq

\end{document}